\documentclass[aps,prb,twocolumn,showpacs,superscriptaddress]{revtex4}

\usepackage{graphicx}
\usepackage{siunitx}
\usepackage{bm}
\usepackage{hyperref}
\usepackage[caption=false]{subfig}
\usepackage{mathtools}
\usepackage{multirow}
\usepackage{amssymb}
\usepackage{braket}
\usepackage{amsmath}
\usepackage{cancel}
\usepackage{xcolor}

\usepackage{spreadtab}

\begin{document}

\title{The temperature dependence of electronic eigenenergies in the adiabatic harmonic approximation}

\author{S. Ponc\'e}
\email{samuel.ponce@uclouvain.be}
\affiliation{%
European Theoretical Spectroscopy Facility, Institute of Condensed Matter and Nanosciences, Universit\'e catholique de Louvain, Chemin des \'etoiles 8, bte L07.03.01, B-1348 Louvain-la-neuve, Belgium.
}%
\author{G. Antonius}%
\affiliation{
D\'epartement de Physique, Universit\'e de Montreal, C.P. 6128, Succursale Centre-Ville, Montreal, Canada H3C 3J7
}%
\author{Y. Gillet}
\affiliation{%
European Theoretical Spectroscopy Facility, Institute of Condensed Matter and Nanosciences, Universit\'e catholique de Louvain, Chemin des \'etoiles 8, bte L07.03.01, B-1348 Louvain-la-neuve, Belgium.
}%
\author{P. Boulanger}
\affiliation{%
Institut N\'eel, 25 avenue des Martyrs, BP 166, 38042 Grenoble cedex 9, France
}%
\author{J. Laflamme Janssen}
\affiliation{%
European Theoretical Spectroscopy Facility, Institute of Condensed Matter and Nanosciences, Universit\'e catholique de Louvain, Chemin des \'etoiles 8, bte L07.03.01, B-1348 Louvain-la-neuve, Belgium.
}%
\author{A. Marini}
\affiliation{%
Consiglio Nazionale delle Ricerche (CNR),Via Salaria Km 29.3, CP 10, 00016, Monterotondo Stazione, Italy
}%
\author{M. C\^ot\'e}
\affiliation{
D\'epartement de Physique, Universit\'e de Montreal, C.P. 6128, Succursale Centre-Ville, Montreal, Canada H3C 3J7
}%
\author{X. Gonze}
\affiliation{%
European Theoretical Spectroscopy Facility, Institute of Condensed Matter and Nanosciences, Universit\'e catholique de Louvain, Chemin des \'etoiles 8, bte L07.03.01, B-1348 Louvain-la-neuve, Belgium.
}%

\date{\today}

\begin{abstract}

The renormalization of electronic eigenenergies due to electron-phonon interactions (temperature dependence and zero-point motion effect) is important in many materials. We address it in the adiabatic harmonic approximation, based on first principles (e.g. Density-Functional Theory), from different points of view: directly from atomic position fluctuations or, alternatively, from  Janak's theorem generalized to the case where the Helmholtz free energy, including the vibrational entropy, is used.  
We prove their equivalence, based on the usual form of Janak's theorem and on the dynamical equation.
We then also place the Allen-Heine-Cardona (AHC) theory of the renormalization in a first-principle context. The AHC theory relies on the rigid-ion approximation, and naturally leads to a self-energy (Fan) contribution and a Debye-Waller contribution. Such a splitting can also be done for the complete harmonic adiabatic expression, in which the rigid-ion approximation is not required.
A numerical study within the Density-Functional Perturbation theory framework allows us to compare the AHC theory with frozen-phonon calculations, with or without the rigid-ion terms.
For the two different numerical approaches without rigid-ion terms, the agreement is better than 7 $\mu$eV in the case of diamond, which represent an agreement to 5 significant digits. The magnitude of the non rigid-ion terms in this case is also presented, distinguishing specific phonon modes contributions to different electronic eigenenergies.

\end{abstract}

\pacs{63.20.kd,65.40.-b,71.15.Mb,71.38.-k}

\maketitle

\section{Introduction}
\label{Introduction}

Electronic properties of solids and nanostructures can be computed from first principles with varying accuracies.
In particular, the widespread $GW$ approximation\cite{Aulbur1999} within many-body perturbation theory 
describes electronic bandgaps with errors in the range of 0.1-0.3 eV with respect to experiment\cite{Shishkin2007}. 
Excitonic effects can be added based on the  Bethe-Salpeter equation (BSE)\cite{Onida2002}.
However, a crucial ingredient is often disregarded: the electron-phonon interaction.
Incidentally, the most advanced self-consistent $GW$ calculations usually lead to overestimated bandgaps\cite{Shishkin2007},
and most of the remaining discrepancy might be due to the electron-phonon interaction.
Indeed, the influence of the lattice vibration at 0~Kelvin, known as the zero-point motion renormalization (ZPR), can be as large as 0.37~eV for the indirect bandgap of diamond\cite{Clark1964,Cardona2005}. This correction often leads to a reduction of the bandgap and hence might be crucial to correct the overestimation of self-consistent $GW$ approximation. 

The study of the direct effects of the electron-phonon interaction on the electronic structure has a long history.
From the fifties to the late eighties, they were investigated and computed in a semi-empirical context.
It was first recognized that the temperature dependence of the electronic energies has two different origins: the volume expansion (studied by Shockley and Bardeen\cite{Shockley1950}) and the electron-phonon interactions at constant volume. 
In fact, the effect of the electron-phonon interaction at constant volume is usually the major contribution, and proves to be
the most difficult to compute from first principles. This contribution is the focus of this paper\footnote{Although first-principle studies of thermal expansion are not very frequent, the formalism is well-established (at least for rather symmetric crystals, see e.g. Rignanese \textit{et al}\cite{Rignanese1996}).}.
In a pioneering work, Fr\"ohlich introduced a model Hamiltonian that includes these interactions\cite{Frohlich1950,Frohlich1952,Frohlich1954}. However, his approach leads to overscreening of phonon frequencies and involves empirical parameters\cite{Leeuwen2004}.


In parallel, the self-energy contribution to electronic eigenenergy renormalization due to thermal vibrations was introduced by Fan\cite{Fan1950,Fan1951}. His theory has no adjustable parameters and is based on the first-order perturbed Hamiltonian. Later, Cohen\cite{Cohen1962} used the Fan self-energy to compute the temperature dependence of the germanium bandgap. Also, ideas from electron diffraction theories based on thermally averaged nuclear potentials lead Anton\v{c}\'{i}k\cite{Antoncik1955} and others\cite{Keffer1968,Walter1970,Kasowski1973} to develop 
empirical Debye-Waller (DW) corrections to the nuclear potential. As these two lines of thought (Fan and DW) were developed independently of each other, only one of them was usually included in calculations.

At about the same time as Anton\v{c}\'{i}k, Brooks\cite{Brooks1955} emphasized that the electronic gap is actually a free energy difference with respect to varying occupation numbers of electronic and phononic levels. 
Furthermore, Allen and Hui\cite{Allen1980} highlighted the equivalence between the action of the phonon population (atomic position fluctuations) on the electronic eigenenergies and the action of the electronic occupations on the phonon eigenfrequencies.  
This equivalence, that was later called Brook's theorem, was used in the eighties by several authors to discuss the temperature dependence of eigenenergies\cite{Allen1980,King-Smith1989}.

In 1974, Baumann\cite{Baumann1974} first suggested that both the Fan self-energy and DW terms were needed to describe the influence of lattice vibrations on the electronic eigenenergies. Two years later, Allen and Heine\cite{Allen1976} rigorously unified the theory and made the DW term translationally invariant. Their approach, combined with the rigid-ion approximation (RIA), which is valid for their semi-empirical model, allows for a re-writing of the problem in terms of first-order derivatives of the effective potential only.  Calculations of electron-phonon renormalization were then led by Cardona and  coworkers\cite{Allen1981,Allen1983,Lautenschlager1985,Zollner1992}, including Allen, based on physically motivated models\cite{Heine1976,Vechten1979} or on rigid-ion pseudopotentials approximations and empirical phonon models. 
The resulting approach is now called the Allen-Heine-Cardona (AHC) theory. 

Until then, none of the calculation were based on first principles. In 1989, King-Smith \textit{et al}\cite{King-Smith1989} computed the temperature-dependent bandgap of Silicon using Density Functional Theory (DFT)\cite{Martin2004}, by evaluating the change 
of phonon frequencies due to electronic occupations and invoking Brook's theorem. 

It took more than one decade before other first-principles calculations were performed.
Such studies relied on widely varying formalisms and methods, that can be broadly classified in three types, each with their distinct advantages and drawbacks. First, the temperature-dependent eigenenergies can be computed as a time average of the band gap obtained using first-principles molecular dynamics (MD) simulations. Using this method, Franceschetti\cite{Franceschetti2007} studied Si nanocrystals, Kamisaka \textit{et al}\cite{Kamisaka2008} studied CdSe and PbSe quantum dots and Ibrahim\cite{Ibrahim2008} the temperature dependence of the optical response of GaAs. Ram\'irez \textit{et al}\cite{Ramirez2006,Ramirez2008} simulated the temperature dependence of diamond and 3C-SiC bandgap based on path-integral molecular dynamics (PIMD). This approach has the interesting characteristics that it includes effects beyond the harmonic approximation. Moreover, while normal molecular dynamics, which involves the harmonic approximation, wrongly delivers a classical Boltzmann statistics for phonons, the more computationally demanding PIMD properly includes nuclear quantum effects and delivers Bose-Einstein statistics for phonons. Thus, with PIMD, zero-point motion effects are observed. 
However, this MD (or PIMD) method is most suited for finite systems. Indeed, for solids, a supercell has to be used to sample the phonon wavevector space, so that the eigenenergies are not well separated from each others and only the band edges can be clearly identified. 

A second method uses frozen phonons (FP): the computation of the change of eigenenergies
due to atomic displacements along the normal modes is followed by a Bose-Einstein weighted sum of the contribution of each mode. In 2005, Capaz \textit{et al}\cite{Capaz2005} studied the temperature dependence of carbon nanotube bandgap within the framework of a tight-binding method. Patrick \textit{ et al}\cite{Patrick2013} examined diamondoids, and Han and Bester \cite{Han2013} studied various semiconductor nanoclusters, still using the FP method but this time with DFT simulation. 
Anharmonic electron-phonon contribution to the temperature dependence of the indirect band gap of diamond were also studied by Monserrat \textit{et al}\cite{Monserrat2013} with the same methodology. Recently, Antonius \textit{ et al}\cite{Antonius2014}, still relying on the FP method, computed the renormalization of the diamond bandgap within the GW approximation and observed a large increase of the renormalization with respect to DFT, in better agreement with experimental values. This result is in line with earlier estimations of many-body effects on electron-phonon coupling\cite{Lazzeri2008,Grueneis2009,LaflammeJanssen2010,Faber2011,Ciuchi2012,Yin2013}.

As a third approach, the diagrammatic method of many-body perturbation theory, from which the AHC approach originates, allowed Giustino \textit{et al} \cite{Giustino2010} to compute the ZPR and the temperature dependence of the diamond bandgap with Wannier functions in the Density-Functional perturbation theory (DFPT)\cite{Baroni1987,Gonze1997} framework. Marini \textit{et al}\cite{Marini2008,Cannuccia2011,Cannuccia2012,Kawai2014} focused on the dynamical effects, beyond the adiabatic approximation, which are absent from the two previous approaches (MD and FP).

There has been some confusion about the (non)-equivalence of these three approaches, in the first-principles context. 
Although the first (MD) and second (FP)  one are equivalent when considered within the adiabatic harmonic approximation, the third one (AHC) 
is equivalent to the MD and FP \textit{only} when the rigid-ion approximation is valid, which is not the case in the first-principle context.
Indeed, Gonze \textit{et al}\cite{Gonze2011} pointed out that the FP and AHC formalisms differ by non-diagonal Debye-Waller terms, and computed these for simple diatomic molecules. In some cases, the non-diagonal Debye-Waller terms were as large as the direct terms. 
They also reformulated the AHC theory using Sternheimer equations instead of summations over empty states, which led to a significant speed-up of their calculations. 
The difference between AHC and FP was also examined in the above-mentioned study by Antonius \textit{et al}\cite{Antonius2014}, where the global effect of the RIA on the ZPR was found to be rather small for the bandgap of diamond.
Recently, a thorough validation study by Ponc\'e \textit{et al}\cite{Ponce2014}, comparing different first-principles codes, 
allowed to resolve a persisting disagreement on the value of the ZPR for the direct bandgap of diamond and established a value of -0.41 eV from AHC formalism on top of DFT\cite{Ponce2014}. The latter study also revealed the very slow convergence with respect to the number of $\mathbf{q}$-points (phonon wavevectors). Such a slow convergence is not restricted only to the AHC method: frozen phonons of many wavevectors should be taken into account for obtaining properly converged FP calculations as well, while huge supercells should be used in the case of MD-based approaches.

To complete this literature overview, the phonon-induced lifetime broadening of the electronic states derived from the imaginary part of the Fan self-energy was investigated by Lautenschlager\cite{Lautenschlager1986,Gopalan1987} in a semi-empirical context, and, more recently, by Giustino\cite{Giustino2007} and Restrepo\cite{Restrepo2009,Restrepo2012} with a first-principle implementation.

In this paper, we will clarify or establish links between the different approaches at the first-principle level, for semiconductors and insulators. Strictly speaking, because of the adiabatic approximation, our theory does not apply to metals, as the phonon frequencies cannot be neglected with respect to electronic excitations. The same limitation is also encountered 
for semiconductors and insulators with a temperature sufficiently high to create a non-negligible population of holes and conduction electrons.  
So, provided the adiabatic approximation is valid, we establish Brook's theorem, 
and provide a detailed analysis of the difference between the AHC approach and the FP approach, 
elaborating on the brief results presented in Ref.~\onlinecite{Gonze2011}. 
Detailed DFT numerical results for diamond will also be provided, 
going, for DFT, further than Ref.~\onlinecite{Antonius2014}.

The structure of the paper is as follows.
Section~\ref{Theory} examines Brooks theorem which relies, at the first-principle level, on Janak's theorem generalized to the case where the free energy including
the vibrational entropy is used, instead of a purely electronic expression. Section~\ref{link_w_AHC} links the equations of section~\ref{Theory} with the AHC theory, in the periodic case (with notations suitable for later practical implementation) and also explores rigorously the rigid-ion approximation. In particular, one distinguishes the Fan contribution, the diagonal rigid-ion Debye-Waller contribution, the diagonal non-rigid-ion Debye-Waller contribution, and the non-diagonal Debye-Waller contribution. The sum over a large number of bands, present in the AHC theory, can be reduced drastically, by complementing it with an expression based on the projection over high energy bands of the first-order wavefunctions, in the spirit of the Sternheimer equation. 
Section~\ref{finite_differences} establishes the connection between those equations and an equivalent finite-difference approach. Then, this connection is used in section~\ref{Calculation_Result} to validate the theory of section~\ref{link_w_AHC} as well as its implementation by comparison with finite-difference calculations for the case of diamond. 
This section also allows one to assess the importance of the rigid-ion approximation in the case of periodic solids. 

All equations derived in this work are expressed in Hartree atomic unit where $m_e=\hbar=e=1$.

\section{Temperature-dependent electronic eigenenergies in the adiabatic harmonic approximation}
\label{Theory}

Brook's ``theorem'' states that the electron-phonon interaction can be regarded either as the shift in an electronic eigenenergy (labeled $n$) when a phonon is added in a phononic mode (labeled $m$) or, equivalently, as the shift in the $m$ phonon mode eigenfrequency when an electron is placed in the $n$ level. Namely, \cite{Brooks1955,Allen1980,King-Smith1989}
\begin{equation}\label{Brookss}
\frac{\partial \varepsilon_n}{\partial n_m} = \frac{\partial \omega_m}{\partial f_n},
\end{equation}
$\varepsilon_n$ and $f_n$ being respectively the eigenenergy and the thermal average of the occupation of the electronic state $n$, while $\omega_m$ and $n_m$ are the eigenfrequency and the thermal average of the phononic occupation of the phonon mode $m$, respectively.

In this section, we detail this equality in a first-principles context: the phonons are obtained in the harmonic adiabatic approximation, from the interatomic force constants, themselves second-order derivatives of the Born-Oppenheimer total energy in which the electronic occupations effects can be traced explicitly.

We decided, for pedagogical reasons, to work in this section with isolated systems but the extension to periodic systems is straightforward using the convention of appendices~\ref{conventionfor} and \ref{perturbationth}.

\subsection{Fluctuations of atomic positions}

The eigenfrequencies $\omega_m$ and unitless mass-scaled eigendisplacements $\xi_{m,\kappa\alpha}$ for the phonon mode $m$ can be obtained from the dynamical equation\cite{Gonze1997a}
\begin{equation}
 \sum_{\kappa\alpha}D_{\substack{\kappa\alpha\\ \kappa'\gamma}}\xi_{m,\kappa\alpha} = \omega_m^2\xi_{m,\kappa'\gamma},
\end{equation}
where $\kappa$ and $\kappa'$ label atoms in the unit cell, $\alpha$ and $\gamma$ label the cartesian spatial dimensions, and $D_{\substack{\kappa\alpha\\ \kappa'\gamma}}$ are the mass-scaled interatomic force constants. 

The mass-scaled quantities are expressed as follows
\begin{equation}
\begin{split}
 \xi_{m,\kappa\alpha}&=\sqrt{M_\kappa}U_{m,\kappa\alpha},\\
 D_{\substack{\kappa\alpha\\ \kappa'\gamma}} &= \frac{1}{\sqrt{M_\kappa}}C_{\substack{\kappa\alpha\\ \kappa'\gamma}}\frac{1}{\sqrt{M_{\kappa'}}},
\end{split}
\end{equation}
where $M_\kappa$ is the mass of atom $\kappa$, $U_{m,\kappa\alpha}$ is the phonon eigendisplacements, and 
\begin{equation}
C_{\substack{\kappa\alpha\\ \kappa'\gamma}}=\frac{\partial^2 E^{BO}}{\partial R_{\kappa\alpha}\partial R_{\kappa'\gamma}},
\end{equation}
are the inter-atomic force constants (IFC), where $E$ is the Born-Oppenheimer  energy of the system (which excludes the kinetic energy of the nuclei), computed from first principles. Using the fact that the mass-scaled eigendisplacements $\xi_{m,\kappa\alpha}$ are orthonormal and complete, the phonon frequencies can be obtained from
\begin{equation}\label{ch2.6}
 \omega_m^2 = \sum_{\substack{\kappa\alpha\\ \kappa'\gamma}}U_{m,\kappa'\gamma}^*C_{\substack{\kappa\alpha\\ \kappa'\gamma}} U_{m,\kappa\alpha}
\end{equation}
and the normalisation of phonon eigendisplacements is
\begin{equation}\label{normalization1}
\sum_{\kappa\alpha}M_{\kappa} U_{m,\kappa\alpha}^* U_{m',\kappa\alpha} =\delta_{mm'}.
\end{equation}

Within the harmonic approximation, where normal modes are decoupled from each others, we express the static temperature dependence of the eigenenergies $\varepsilon_n(T)$ as a thermal average of the 
value of the temperature-independent, position-dependent eigenenergies $\tilde{\varepsilon}_n[z\mathbf{U}]$, where $z\mathbf{U}$ denotes generically a displacement from the equilibrium atomic positions
\begin{align}\label{eqdeptempisole}
\varepsilon_n(T)\triangleq & \left\langle \tilde{\varepsilon}_n[z\mathbf{U}]\right\rangle (T)\\
=&\sum_m^{3N} \frac{1}{\mathcal{Z}_m} \sum_{s_m} e^{\frac{-(s_m+1/2)\omega_m}{k_B T}}\nonumber\\
 & \qquad \quad  \int \chi_{m,s_m}^*(z) \tilde{\varepsilon}_n[z \mathbf{U}_{m,\kappa}]\chi_{m,s_m}(z)dz,
\end{align}
with $N$ the number of atom, $T$ the temperature, $s_m$ the integer occupation of phonon mode $m$, $\mathcal{Z}_m=\sum_{s_m} e^{\frac{-(s_m+1/2)\omega_m}{k_B T}}$ the mode-partition function, $k_B$ the Boltzmann's constant, $\chi_{m,s_m}(z)$ the phonon eigenfunctions, $z$ the spatial coordinate associated with a phonon mode and where bold symbols like $\mathbf{U}_{m,\kappa}$ denote cartesian vectors. 

We can expand the eigenenergies $\tilde{\varepsilon}_n[z \mathbf{U}_{m,\kappa}]$ of Eq. \eqref{eqdeptempisole} in a Taylor series up to second order in $z$, since we are working within the harmonic approximation
\begin{multline}
\tilde{\varepsilon}_n[z \mathbf{U}_{m,\kappa}] = \tilde{\varepsilon}_n[0]+z\frac{d}{dz}\tilde{\varepsilon}_n[z \mathbf{U}_{m,\kappa}]\Big|_{z=0}\\
+\frac{1}{2}z^2 \frac{d^2}{dz^2}\tilde{\varepsilon}_n[z \mathbf{U}_{m,\kappa}]\Big|_{z=0},
\end{multline}
and insert it in Eq. \eqref{eqdeptempisole}. Using the properties of phononic wavefunctions in an harmonic potential (see appendix~\ref{integral} for more details), we obtain 
\begin{multline}
\varepsilon_n(T) = \tilde{\varepsilon}_n[0] + \frac{1}{2}\sum_m^{3N} \frac{1}{\mathcal{Z}_m \omega_m} \frac{d^2}{dz^2}\tilde{\varepsilon}_n[z \mathbf{U}_{m,\kappa}]\Big|_{z=0}\\
\sum_{s_m} e^{-(s_m+1/2)\alpha_m}\left(\frac{2s_m+1}{2}\right)
\end{multline}
where we have defined $\alpha_m \triangleq \frac{\omega_m}{k_BT}$.

For sake of brevity, we introduce the notation
$\Delta \varepsilon_{n}(T)  \triangleq \varepsilon_{n}(T) - \tilde{\varepsilon}_{n}[0]$.
Using the properties of geometrical series to evaluate the sum over $s_m$, we get 
\begin{multline}
\Delta\varepsilon_n(T) = \frac{1}{2}\sum_m^{3N} \frac{1}{\frac{e^{-\alpha_m/2}}{1-e^{-\alpha_m}}\omega_m} \frac{d^2}{dz^2}\tilde{\varepsilon}_n[z \mathbf{U}_{m,\kappa}]\Big|_{z=0}\\
 e^{-\alpha_m/2}\left( \frac{e^{-\alpha_m}}{(1-e^{-\alpha_m})^2} + \frac{1}{2(1-e^{-\alpha_m})} \right),
\end{multline}
which simplifies to
\begin{multline}\label{firstequivalent}
\Delta\varepsilon_n(T)  = \sum_m^{3N} \frac{1}{2\omega_m} \frac{d^2}{dz^2}\tilde{\varepsilon}_n[z \mathbf{U}_{m,\kappa}]\Big|_{z=0}\\
 \left( n_m(T) + \frac{1}{2}\right),
\end{multline}
where we have introduced the Bose-Einstein distribution 
\begin{equation}\label{bose-einstein}
n(T)=\frac{1}{e^{\frac{\omega_m}{k_B T}}-1}.
\end{equation}

We can consider the phonon occupation numbers $n_m(T)$ as independent variables in this expression, in which case
\begin{equation}\label{Brookss2}
\frac{\partial \varepsilon_n}{\partial n_m}\triangleq \frac{\partial \varepsilon_n(T)}{\partial n_m(T)}=
\frac{1}{2\omega_m} \frac{d^2}{dz^2}\tilde{\varepsilon}_n[z \mathbf{U}_{m,\kappa}]\Big|_{z=0},
\end{equation}
where we have defined the short-hand notation on the left hand side of the equation because each term in the central part of the equation depends on the temperature but their ratio does not. The slope of the eigenenergies with the phononic occupation is therefore independent of temperature and we will remove the explicit dependence of the terms when considering the slope. Eq.~\eqref{firstequivalent} can thus be written
\begin{equation}\label{Eq.12}
\Delta\varepsilon_n(T)  = \sum_m^{3N} \frac{\partial \varepsilon_n}{\partial n_m} \left( n_m(T) + \frac{1}{2}\right).
\end{equation}
Then, the zero-point motion contribution (ZPR) is simply the shift of the eigenenergies
  at $T=0$, that is
\begin{equation}
\Delta\varepsilon_n(T=0)  =  \frac{1}{2}\sum_m^{3N} \frac{\partial \varepsilon_n}{\partial n_m},
\end{equation}
 highlighting that $\varepsilon_n(T=0)\neq \varepsilon_n[0]$.
Additionally, from Eqs.~\eqref{ch2.6} and \eqref{normalization1} we can see that the phonon frequency goes as $M_\kappa^{-1/2}$.
The $\mathbf{U}_{m,\kappa}$ goes as $M_\kappa^{-1/2}$ and the $z\mathbf{U}_{m,\kappa}$ must have the dimension of length because it is an ionic displacement (see Eq.~\eqref{firstequivalent}). Therefore, the $z$ must have the dimension of $M_\kappa^{1/2}$ times length. All of this leads to the fact that the ZPR of Eq.~\eqref{firstequivalent} goes as $(\omega_m M_\kappa)^{-1}$, i.e. as $M_\kappa^{-1/2}$. This ``isotopic effect'' allows for experimental measurements of the zero-point renormalization by substituting atoms with heavier isotopes, as explained in the review paper of Cardona and Thewalt~\cite{Cardona2005}.

\subsection{Eigenenergies as derivatives with respect to electronic occupation numbers $\mathbf{f_n}$}

Following  Janak, we extend to fractional occupations the first-principle Born-Oppenheimer energy $E^{BO}$.  Janak's theorem\cite{Janak1978} then gives
\begin{equation}\label{Janakeqperiodic}
\varepsilon_{n}=\frac{\partial E^{BO}}{\partial f_{n}},
\end{equation}
where it has to be noted that the Janak theorem breaks down, within DFT, if the exact exchange-correlation functional is used. In that case, the total energy is not a continuous function of the electronic occupation anymore and eigenenergies must be defined as difference of total energies with integral occupations numbers.

We now complement the Born-Oppenheimer energy with phonon energy and entropy at the harmonic level. The energy becomes
\begin{equation}\label{eqE}
\begin{split}
 E(T) &=  E^{BO} + E_{\text{vib}}(T)\\
      &= E^{BO} + \sum_{m}^{3N}\omega_m\left( n_m(T)+\frac{1}{2}\right),
\end{split}
\end{equation}
where $E^{BO}$ is the total energy without phonon and electron-phonon contributions. Taking into account the vibrational entropy gives a Helmholtz free energy\cite{Fultz2010}
\begin{equation}\label{eqfreenergy}
F(T) \triangleq  E(T) - TS_{\text{vib}}(T),
\end{equation}
where $S_{\text{vib}}$ is the vibrational entropy,
\begin{multline}\label{ch2.26}
  S_{\text{vib}}(T)=k_B \sum_m^{3N}\Big((1+n_m(T))\ln(1+n_m(T)) \\
  - n_m(T)\ln(n_m(T))\Big).
\end{multline}

We now show that the temperature-dependent eigenenergies can be obtained from the extension of Janak's theorem to finite phonon temperature
\begin{equation}\label{Janaktemperature}
\varepsilon_{n}(T)=\frac{\partial F(T)}{\partial f_{n}}.
\end{equation}
For sake of simplicity, we neglect the dependence of 
the electronic occupations $f_n$ on electronic temperature. Actually, the
explicit treatment of the electron system at finite temperature (e.g. using the Mermin functional\cite{Mermin1965}), 
supposing (wrongly) the adiabatic approximation to be still valid, would not change the remaining of the paper.
Using this definition of $\varepsilon_{n}(T)$, Eq.~\eqref{eqE}, Eq.~\eqref{eqfreenergy} and taking into account the dependence of the phonon frequencies on electronic occupation numbers as well as the dependence of phonon occupation numbers on electronic occupation numbers, the change of eigenenergies due to electron-phonon interaction becomes
\begin{equation}\label{Janakderivation}
\begin{split}
\Delta \varepsilon_{n}(T)  
          =& \sum_{m}^{3N} \bigg( \frac{\partial \omega_m}{\partial f_{n}} \Big(  n_m(T)+\frac{1}{2}\Big)\\
 &+ \omega_m\frac{\partial n_m(T)}{\partial f_{n}}-T\frac{\partial S_{\text{vib}}}{\partial n_m(T)}\frac{\partial n_m(T)}{\partial f_{n}}\bigg).
\end{split}
\end{equation} 

Substituting Eq. \eqref{ch2.26} for $S_{\text{vib}}$ into Eq. \eqref{Janakderivation} gives
\begin{multline}
\Delta \varepsilon_{n}(T)=\sum_{m}^{3N} \left( \frac{\partial \omega_m}{\partial f_{n}} \left(  n_m(T)+\frac{1}{2}\right)\right.+ \omega_m\frac{\partial n_m(T)}{\partial f_{n}}\\
-\left.k_B T\frac{\partial n_m(T)}{\partial f_{n}}\ln\left( \frac{1+n_m(T)}{ n_m(T)} \right)\right).
\end{multline}
Using Eq.~\eqref{bose-einstein}, the last 2 terms in the sum cancel out. We thus obtain 
\begin{equation}\label{ch2.7bis}
\Delta \varepsilon_{n}(T) = \sum_{m}^{3N}   \frac{\partial \omega_m}{\partial f_{n}} \left(  n_m(T)+\frac{1}{2}\right).
\end{equation}

This is a first important result of the present paper. To the authors knowledge, it was never derived starting from the free energy Eq.~\eqref{eqfreenergy} and the finite temperature extension of Janak's theorem Eq.~\eqref{Janaktemperature}. 
Identification of Eq.~\eqref{ch2.7bis} with Eq.~\eqref{Eq.12} obviously yields
\begin{equation}\label{secondversion}
\frac{\partial \varepsilon_{n}}{\partial n_m} = \frac{\partial \omega_m}{\partial f_{n}}.
\end{equation}

This link can be more rigorously established as follows. The derivative of the phonon frequency is retrieved from deriving Eq. \eqref{ch2.6} with respect to electronic occupation
\begin{equation}\label{phononfreq}
 2\omega_m\frac{\partial \omega_m}{\partial f_{n}} = \sum_{\substack{\kappa\alpha\\ \kappa'\gamma}} \frac{\partial C_{\substack{\kappa\alpha\\ \kappa'\gamma}}}{\partial f_{n}}  U_{m,\kappa'\gamma}^*U_{m,\kappa\alpha},
\end{equation} 
where the derivative of the displacements $U_{m,\kappa\alpha}$ with respect to the occupations $f_n$ does not contribute thanks to the Hellmann-Feynman theorem\cite{Hellmann1937,Feynman1939}. 
The temperature dependence of the eigenenergies is obtained by substituting the preceding result for $\frac{\partial \omega_m}{\partial f_n}$ into Eq.~\eqref{ch2.7bis} 
\begin{multline}\label{ch2.17}
 \Delta\varepsilon_{n}(T)=   \sum_{m}^{3N} \frac{1}{2\omega_m}
 \sum_{\substack{\kappa\alpha\\ \kappa'\gamma}}U_{m,\kappa'\gamma}^* \\
\frac{\partial C_{\substack{\kappa\alpha\\ \kappa'\gamma}}}{\partial f_{n}}
 U_{m,\kappa\alpha}\left(  n_m(T)+\frac{1}{2}\right).
\end{multline}
Using Janak's theorem, we can reformulate the derivative of the IFC into derivatives of eigenenergies
\begin{equation}\label{eqJanakIFC}
 \frac{\partial C_{\substack{\kappa\alpha\\ \kappa'\gamma}} }{\partial f_{n}}= \frac{\partial}{\partial f_{n}}\left(\frac{\partial^2 E^{BO}}{\partial R_{\kappa\alpha}\partial R_{\kappa'\gamma}}\right) = \frac{\partial^2 \varepsilon_{n}}{\partial R_{\kappa\alpha}\partial R_{\kappa'\gamma}}.
\end{equation}
Substituting the above expression into Eq.~\eqref{ch2.17}, we obtain
\begin{multline}\label{eqtempdepisolated}
 \Delta\varepsilon_{n}(T)=   \sum_{m}^{3N} \frac{1}{2\omega_m}
 \sum_{\substack{\kappa\alpha\\ \kappa'\gamma}}U_{m,\kappa'\gamma}^* \\
\frac{\partial^2 \varepsilon_{n}}{\partial R_{\kappa\alpha}\partial R_{\kappa'\gamma}}
 U_{m,\kappa\alpha}\left(  n_m(T)+\frac{1}{2}\right).
\end{multline}
The double sum over atomic position displacements is actually the second-order derivative of the eigenenergy
with respect to the normal mode,
\begin{multline}\label{eqsecondeigen}
 \sum_{\substack{\kappa\alpha\\ \kappa'\gamma}}U_{m,\kappa'\gamma}^*
\frac{\partial^2 \varepsilon_{n}}{\partial R_{\kappa\alpha}\partial R_{\kappa'\gamma}}
 U_{m,\kappa\alpha}
 =\\
 \frac{d^2}{dz^2} \tilde{\varepsilon}_n[z \mathbf{U}_{m,\kappa}]\Big|_{z=0},
\end{multline}
so that we recover, from Eqs.~\eqref{Brookss2}, \eqref{secondversion}  and \eqref{eqsecondeigen}  the following relations
\begin{equation}\label{equallity}
\begin{split}
\frac{\partial \varepsilon_n}{\partial n_m}=& \frac{1}{2\omega_m}\frac{d^2}{dz^2} \tilde{\varepsilon}_n[z \mathbf{U}_{m,\kappa}]\Big|_{z=0}\\
=&\frac{1}{2\omega_m}\sum_{\substack{\kappa\alpha\\ \kappa'\gamma}}U_{m,\kappa'\gamma}^*
\frac{\partial^2 \varepsilon_{n}}{\partial R_{\kappa\alpha}\partial R_{\kappa'\gamma}} U_{m,\kappa\alpha} = \frac{\partial \omega_m}{\partial f_n}.
\end{split}
\end{equation}
We thus obtain a demonstration of the Brook's theorem in the first-principle context.

\subsection{The Fan and Debye-Waller contributions}

In order to establish the links with the Fan, Debye-Waller and AHC approaches, we now analyze Eq. \eqref{eqtempdepisolated} in more detail, focusing on the second-order derivative of the eigenenergies with respect to two atomic displacements. We can obtain it from perturbation theory. We start from
\begin{equation}\label{eigenenergy}
\varepsilon_n = \Bra{\Psi_{n}}\hat{H}\Ket{\Psi_{n}},
\end{equation}
and differentiate it, also using the Hellmann-Feynman theorem. At the equilibrium geometry,
\begin{equation}
\begin{split}\label{ch2.8bis}
\frac{\partial \varepsilon_n}{\partial R_{\kappa\alpha}} =  \Bra{\Psi_n^{(0)}} \frac{\partial\hat{H}}{R_{\kappa\alpha}}\Ket{ \Psi_n^{(0)}}.
\end{split}
\end{equation}

Eq.~\eqref{ch2.8bis} can be derived a second time, with respect to another atomic displacement. An equivalent result can be obtained by switching the two atomic displacements. Both can be combined and deliver an expression that is real and explicitly symmetric with respect to the indices $\kappa\alpha$ and $\kappa'\gamma$
\begin{multline}\label{eqsolidisoleE}
\frac{\partial^2 \varepsilon_{n}}{\partial R_{\kappa\alpha}\partial R_{\kappa'\gamma}} = \Bra{\Psi_{n}^{(0)}}\frac{\partial^2\hat{H}}{\partial R_{\kappa\alpha}R_{\kappa'\gamma}} \Ket{\Psi_{n}^{(0)}} + \\
 \frac{1}{2}\bigg[\Big( 
\Big\langle\frac{\partial\Psi_{n}}{\partial R_{\kappa\alpha}} \Big|
 \frac{\partial \hat{H}}{\partial R_{\kappa'\gamma}} \Ket{\Psi_{n}^{(0)}} + (\kappa\alpha) \leftrightarrow (\kappa'\gamma)\Big)\\
+(c.c.)\bigg], 
\end{multline}
where $(\kappa\alpha) \leftrightarrow (\kappa'\gamma)$ stands for the previous term in which the indices $\kappa\alpha$ and $\kappa'\gamma$ have been exchanged, and where $(c.c.)$ stands for the complex conjugate of the previous term. 

The contribution from the second-order perturbation of the Hamiltonian, $\frac{\partial^2\hat{H}}{\partial R_{\kappa\alpha}R_{\kappa'\gamma}}$, gives the Debye-Waller (DW) term\cite{Antoncik1955,Yu1964} of the semi-empirical approach.  
The other bracketed term originates from the first-order modifications of the wavefunction and corresponds to the Fan term when
considered in many-body perturbation theory. As mentioned in the introduction, the complementarity of the two terms for the description of the eigenenergy renormalization due to the electron-phonon interaction, although obvious in the
present derivation, was first shown in 1974 by Baumann \cite{Baumann1974}.

\section{From the adiabatic approximation to the Allen-Heine-Cardona theory}
\label{link_w_AHC}

In this section, we relate Eq.~\eqref{eqtempdepisolated} to the AHC theory, as formulated\footnote{Although we agree with Eqs.~(2-7) of Allen and Cardona\cite{Allen1981}, we think that their Eq.~(8) is erroneous, since the polarization vectors in the Debye-Waller term of this equation have different atomic indices $\kappa$ and $\kappa'$, which is not coherent with their Eq.~(5) or with Ref.~\onlinecite{Giustino2010}, where the polarization vectors have the same atomic index $\kappa$. On the contrary, we perfectly agree with the formulation of Giustino \textit{ et al.}\cite{Giustino2010}.} by Giustino \textit{ et al.}\cite{Giustino2010}.
 We also carefully treat the RIA and unveil the terms it neglects. Furthermore, we discuss the translational invariance and its consequences. In order to focus on solids (in view of the application and testing for diamond and solid-state systems in general), we work with periodic systems and introduce the related definitions. More information in this respect is provided in appendix~\ref{conventionfor} and \ref{perturbationth}.

\subsection{Wavevectors and translational invariance}

We introduce a specific notation for the derivative of an arbitrary quantity $X$, that depends on the atomic coordinates, with respect to a collective displacement of atoms characterized by a wavevector $\mathbf{q}$
\begin{equation}\label{defbasederiv}
\frac{\partial X}{\partial R_{\kappa\alpha}(\mathbf{q})} = \frac{1}{N_{BvK}}\sum_l e^{i\mathbf{q}\cdot \mathbf{R}_l}\frac{\partial X}{\partial R_{l\kappa\alpha}},
\end{equation}
where $R_{l\kappa\alpha}$ is the coordinate along the $\alpha$ axis of the atom $\kappa$ in the cell $l$,
and $N_{BvK}$ is the number of primitive cells of the periodic system defined by the Born-von Karman boundary conditions\cite{Born1954}.

The eigendisplacement vectors are solution of the dynamical equation
\begin{equation}
\sum_{\kappa\alpha}\tilde{C}_{\substack{\kappa\alpha\\ \kappa'\gamma}}(\mathbf{q})U_{m,\kappa\alpha}(\mathbf{q}) = M_{\kappa'}\omega_{m\mathbf{q}}^2U_{m,\kappa'\gamma}(\mathbf{q}),
\end{equation}
where the Fourier transform of the IFC appears 
\begin{equation}\label{ch2.16}
\begin{split}
 \tilde{C}_{\substack{\kappa\alpha\\ \kappa'\gamma}}(\mathbf{q}) =& \frac{1}{N_{BvK}}\sum_{ll'} C_{\substack{l\kappa\alpha\\ l'\kappa'\gamma}} e^{-i\mathbf{q}\cdot(\mathbf{R}_l-\mathbf{R}_{l'})}\\
              =& \sum_{l'} C_{\substack{0\kappa\alpha\\ l'\kappa'\gamma}} e^{i\mathbf{q}\cdot\mathbf{R}_{l'}}\\
=&N_{BvK}\frac{\partial^2 E^{BO}}{\partial R_{\kappa'\alpha}(\mathbf{-\mathbf{q}})\partial R_{\kappa''\beta}(\mathbf{\mathbf{q}})},
\end{split}
\end{equation}
where $\mathbf{R}_{l}$ is a translation vector of the Bravais lattice.
The eigendisplacement vectors fulfill the following normalization condition
\begin{equation}
\sum_{\kappa\alpha}M_{\kappa}U_{m,\kappa\alpha}^*(\mathbf{q})U_{m',\kappa\alpha}(\mathbf{q}) = \delta_{mm'}.
\end{equation}

All properties of a crystal, including its eigenenergies and their derivatives, must be invariant upon a uniform translation $\bm{\delta}$. Therefore,
\begin{align}
 \varepsilon_{n\mathbf{k}}[\{\mathbf{R}_{l\kappa}\}] &= \varepsilon_{n\mathbf{k}}[\{\mathbf{R}_{l\kappa}+\bm{\delta}\}] \\
 \frac{\partial \varepsilon_{n\mathbf{k}}}{\partial R_{\kappa'\alpha}(\mathbf{\Gamma})}[\{\mathbf{R}_{l\kappa}\}] &= \frac{\partial \varepsilon_{n\mathbf{k}} }{\partial R_{\kappa'\alpha}(\mathbf{\Gamma})}[\{\mathbf{R}_{l\kappa}+ \bm{\delta}\}].\label{ch2.1bis}
\end{align}   
In Eq.~\ref{ch2.1bis}, we have taken the derivative with respect to a collective displacement of atoms that does not break the translation symmetry
$(\mathbf{q}=\mathbf{\Gamma})$ to avoid any problem with the Bloch theorem, and keep the Bloch notation $n\mathbf{k}$.

By Taylor expanding the right hand side of Eq.~\eqref{ch2.1bis}, we obtain
\begin{multline}
 \frac{\partial \varepsilon_{n\mathbf{k}}}{\partial R_{\kappa'\alpha}(\mathbf{\Gamma})}[\{\mathbf{R}_{l\kappa}\}] = \frac{\partial \varepsilon_{n\mathbf{k}}}{\partial R_{\kappa'\alpha}(\mathbf{\Gamma})}[\{\mathbf{R}_{l\kappa}\}] \\
 + \sum_{\kappa''\beta } \delta_\beta \frac{\partial^2 \varepsilon_{n\mathbf{k}}}{\partial R_{\kappa'\alpha}(\mathbf{\Gamma})\partial R_{\kappa''\beta}(\mathbf{\Gamma})}[\{\mathbf{R}_{l\kappa}\}] + \mathcal{O}(\bm{\delta}^2),
\end{multline}
where $\delta_\beta$ is the $\beta$ component of the vector $\bm{\delta}$.
For the equality to hold for all $\mathbf{\delta}$, every term of order one and higher in the series  must be identically zero  
\begin{align}\label{ch2.18bis}
 \sum_{\beta}\delta_\beta \sum_{\kappa''}\frac{\partial^2 \varepsilon_{n\mathbf{k}}}{\partial R_{\kappa'\alpha}(\mathbf{\Gamma})\partial R_{\kappa''\beta}(\mathbf{\Gamma})}[\{\mathbf{R}_{l\kappa}\}] =& 0 \quad \forall\, \delta_\beta \in \mathbb{R} \nonumber \\
\Rightarrow \sum_{\kappa''}\frac{\partial^2 \varepsilon_{n\mathbf{k}}}{\partial R_{\kappa'\alpha}(\mathbf{\Gamma})\partial R_{\kappa''\beta}(\mathbf{\Gamma})}[\{\mathbf{R}_{\kappa}^l\}] =& 0.
\end{align}

\subsection{The temperature dependence in the adiabatic harmonic approximation for the solid periodic case}

Eq.~\eqref{eqtempdepisolated} can be generalized to the periodic case, with a discretized integral over $\mathbf{q}$
($N_q$ is the number of wavevectors used to sample the Brillouin zone)
\begin{multline}
 \Delta\varepsilon_{n\mathbf{k}}(T) = \frac{1}{N_q} \sum_{\mathbf{q}} \sum_{m}^{3N} \frac{1}{2\omega_{m\mathbf{q}}}
 \sum_{\substack{\kappa\alpha\\ \kappa'\gamma}}
\sum_{ll'}\frac{\partial^2 \varepsilon_{n\mathbf{k}}}{\partial R_{l\kappa\alpha}\partial R_{l'\kappa'\gamma}}\\
 e^{-i\mathbf{q}\cdot (\mathbf{R}_{l}-\mathbf{R}_{l'})}
U_{m,\kappa'\gamma}^*(\mathbf{q})U_{m,\kappa\alpha}(\mathbf{q})\Big( n_{m\mathbf{q}}(T) +\frac{1}{2}\Big). 
\end{multline}
Similarly, the generalization of Eq.~\eqref{Eq.12} leads to  
\begin{equation}\label{Denk(T)}
 \Delta\varepsilon_{n\mathbf{k}}(T) = \frac{1}{N_q}\sum_{\mathbf{q}} \sum_{m}^{3N} \frac{\partial \varepsilon_{n\mathbf{k}}}{\partial n_{m\mathbf{q}}}\Big(n_{m\mathbf{q}}(T)+\frac{1}{2}\Big),
\end{equation}
where
\begin{multline}\label{denk_dnmq}
 \frac{\partial \varepsilon_{n\mathbf{k}}}{\partial n_{m\mathbf{q}}} = 
    \frac{1}{2\omega_{m\mathbf{q}}}  \sum_{\substack{\kappa\alpha\\ \kappa'\gamma}}
 \sum_{ll'}\frac{\partial^2 \varepsilon_{n\mathbf{k}}}{\partial R_{l\kappa\alpha}\partial R_{l'\kappa'\gamma}}\\
 e^{-i\mathbf{q}\cdot (\mathbf{R}_{l}-\mathbf{R}_{l'})}
U_{m,\kappa'\gamma}^*(\mathbf{q})U_{m,\kappa\alpha}(\mathbf{q})
\end{multline} 
and we will focus on the latter quantity, which represents the change of eigenenergy due to a specific phonon mode.

To split this expression in a Fan and a Debye-Waller contribution, 
we substitute the extension to periodic system of Eq.~\eqref{eqsolidisoleE} in it and retrieve
\begin{multline}\label{zprequation}
 \frac{\partial \varepsilon_{n\mathbf{k}}}{\partial n_{m\mathbf{q}}} = 
 \frac{1}{2\omega_{m\mathbf{q}}}
 \sum_{\substack{\kappa\alpha\\ \kappa'\gamma}} U_{m,\kappa'\gamma}^*(\mathbf{q})U_{m,\kappa\alpha}(\mathbf{q}) \\
\bigg\{ \Bra{ u_{n\mathbf{k}}^{(0)}} \frac{\partial^2 \hat{H}_{\mathbf{k,k}}}{\partial R_{\kappa\alpha}(-\mathbf{q}) \partial R_{\kappa'\gamma}(\mathbf{q})} \Ket{ u_{n\mathbf{k}}^{(0)}} \\
  + \frac{1}{2} \bigg( \Big( \Big\langle \frac{\partial u_{n\mathbf{k}}}{\partial R_{\kappa\alpha}(\mathbf{q})} \Big|\frac{\partial \hat{H}_{\mathbf{k,k}}}{\partial R_{\kappa'\gamma}(\mathbf{q})} \Big|u_{n\mathbf{k}}^{(0)}\Big\rangle \\
   + (\kappa\alpha)\leftrightarrow  (\kappa'\gamma)\Big) + (c.c.)\bigg)\bigg\},
\end{multline}
where $\hat{H}_{\mathbf{k,k}}$ is defined through Eq.~\eqref{def_operator} applied to the Hamiltonian.  

This allows us to introduce the following notation for the DW and Fan contributions related to band $n$ and wavevector $\mathbf{k}$ (we skip the $n$ and $\mathbf{k}$ indices, which should not be confusing in the present context)
\begin{equation}\label{DW}
\mathcal{D}_{\substack{\kappa\alpha\\ \kappa'\gamma}}(\mathbf{q}) \triangleq \Big\langle u_{n\mathbf{k}}^{(0)} \Big| \frac{\partial^2 \hat{H}_{\mathbf{k,k}}}{\partial R_{\kappa\alpha}(-\mathbf{q})\partial R_{\kappa'\gamma}(\mathbf{q})}\Big| u_{n\mathbf{k}}^{(0)}\Big\rangle
\end{equation}
\begin{equation}\label{Fan}
\begin{split}
\mathcal{F}_{\substack{\kappa\alpha\\ \kappa'\gamma}}(\mathbf{q}) \triangleq 
\frac{1}{2} \bigg[ \Big( 
\Big\langle \frac{\partial u_{n\mathbf{k}}}{\partial R_{\kappa\alpha}(\mathbf{q})} \Big| \frac{\partial \hat{H}_{\mathbf{k,k}}}{\partial R_{\kappa'\gamma}(\mathbf{q}) }\Big| u_{n\mathbf{k}}^{(0)}\Big\rangle \\
 + (\kappa\alpha) \leftrightarrow (\kappa'\gamma) \Big) + (c.c.) \bigg].
\end{split}
\end{equation}
The change of eigenenergy due to a specific phonon mode, Eq. \eqref{zprequation}, thus becomes
\begin{multline}\label{denk_dnmq_abbrev}
 \frac{\partial \varepsilon_{n\mathbf{k}}}{\partial n_{m\mathbf{q}}} = 
    \frac{1}{2\omega_{m\mathbf{q}}}  \sum_{\substack{\kappa\alpha\\ \kappa'\gamma}}\\
    \Big[ \mathcal{D}_{\substack{\kappa\alpha\\ \kappa'\gamma}}(\mathbf{q}) 
+ \mathcal{F}_{\substack{\kappa\alpha\\ \kappa'\gamma}}(\mathbf{q}) \Big] U_{m,\kappa'\gamma}^*(\mathbf{q})U_{m,\kappa\alpha}(\mathbf{q}).
\end{multline} 

With the same notations, the translational invariance Eq.~\eqref{ch2.18bis} reads
\begin{equation}\label{TI}
\sum_{\kappa'}\mathcal{D}_{\substack{\kappa\alpha\\ \kappa'\gamma}}(\mathbf{\Gamma})+\mathcal{F}_{\substack{\kappa\alpha\\ \kappa'\gamma}}(\mathbf{\Gamma}) = 0. \qquad \forall\, \alpha, \gamma, \kappa
\end{equation}

In particular, one can multiply this expression by any expression independent of  $\kappa'$,
  and still get zero. This gives us some freedom on the form of the added terms which are chosen in such a way that the $ \mathcal{D}$ terms will cancel out in the rigid-ion approximation (see next subsection). Moreover, we can also 
sum over $\kappa$ instead of $\kappa'$ in Eq.~\eqref{TI} in such a way that the resulting expression is Hermitian
\begin{multline}
\label{denk_dnmq_RI}
 \frac{\partial \varepsilon_{n\mathbf{k}}}{\partial n_{m\mathbf{q}}} = 
    \frac{1}{2\omega_{m\mathbf{q}}}  \sum_{\substack{\kappa\alpha\\ \kappa'\gamma}} \bigg[
        \Big[ \mathcal{D}_{\substack{\kappa\alpha\\ \kappa'\gamma}}(\mathbf{q}) 
          + \mathcal{F}_{\substack{\kappa\alpha\\ \kappa'\gamma}}(\mathbf{q}) \Big]\\
          U_{m,\kappa'\gamma}^*(\mathbf{q})U_{m,\kappa\alpha}(\mathbf{q})
      - \Big[ \mathcal{D}_{\substack{\kappa\alpha\\ \kappa'\gamma}}(\mathbf{\Gamma})
          + \mathcal{F}_{\substack{\kappa\alpha\\ \kappa'\gamma}}(\mathbf{\Gamma}) \Big]\\
          \frac{1}{2}\Big(    U_{m,\kappa\gamma}^*(\mathbf{q})U_{m,\kappa\alpha}(\mathbf{q}) 
          + U_{m,\kappa'\gamma}^*(\mathbf{q})U_{\kappa'\alpha}^m(\mathbf{q})
          \Big)
      \bigg].
\end{multline} 

Beyond the adiabatic and harmonic approximations, we have not made any additional approximation until now.
The Debye-Waller term that we have obtained, Eq.~\eqref{DW}, invokes the second-order derivative of the  first-principle Hamiltonian.
In most DFPT procedures, calculations of the phonon band structure rely only on the evaluation of the
\textit{ first-order} derivative of the self-consistent DFT Hamiltonian and wavefunctions. 
Indeed, while the expression leading to such phonon band structure calculations include a second-order derivative with respect to the non-self-consistent electron-ion potential, the latter term does not depend on \textit{ second-order} derivative of the wavefunction and Hamiltonian.
Thus, the Debye-Waller term is not a by-product of a phonon band structure calculation. 
We will now show how the rigid-ion approximation allows us to compute the Debye-Waller term
without computing the second-order derivative of the  self-consistent first-principle Hamiltonian.
 
\subsection{The rigid-ion approximation}

In the case of semi-empirical potentials, it is natural to suppose that the Hamiltonian 
depend on potentials created \textit{ independently} by each nucleus, screened by electrons attached to them.
In this case, as pointed by Allen and Heine\cite{Allen1976}, the numerical burden of computing a second-order derivative of the Hamiltonian can be completely avoided.
A rigid-ion Hamiltonian has the following form
\begin{equation}\label{rigid-ion_H}
\hat{H}_{\text{ri}} = \hat{T}+\sum_{l\kappa} V_\kappa(\mathbf{\hat{r}}-\mathbf{R}_{l\kappa}).
\end{equation}
so that its mixed (off-site) second-order derivatives vanish
\begin{equation}
\frac{\partial^2 \hat{H}_{\text{ri}}}{\partial R_{l\kappa\alpha}\partial R_{l'\kappa'\gamma}} = 0 \qquad \text{ if } \kappa\neq \kappa' \text{as well as if } l \neq l'.
\end{equation}

Within DFT, imposing such properties amounts to neglecting the effect of the density variation due to the displacement of one atom on the screening of the potential created by the displacement of another atom. 
Using the notation of Eq.~\eqref{DW}, the rigid-ion approximation implies that $\mathcal{D}_{\substack{\kappa\alpha\\ \kappa'\gamma}}(\mathbf{q})$ in RIA is
\begin{equation}\label{NDDWvanish}
\mathcal{D}^{RIA}_{\substack{\kappa\alpha\\ \kappa'\gamma}}(\mathbf{q}) =
\mathcal{D}^{RIA}_{\substack{\kappa\alpha\\ \kappa\gamma}}(\mathbf{q}) \delta_{\kappa, \kappa'} =
\mathcal{D}^{RIA}_{\substack{\kappa\alpha\\ \kappa\gamma}}(\mathbf{\Gamma}) \delta_{\kappa, \kappa'}
\end{equation}
i.e. the non-site-diagonal Debye-Waller contributions vanish.

If we now apply Eq.~\eqref{NDDWvanish} to Eq. \eqref{denk_dnmq_RI}, 
even if the latter is derived from a DFT Hamiltonian, only two terms remain, 
which we call the Fan term and the diagonal Debye-Waller term in the rigid-ion approximation
\begin{equation}\label{Eqrewritednotfinal}
\frac{\partial \varepsilon_{n\mathbf{k}}^{RIA}}{\partial n_{m\mathbf{q}}} =   \frac{\partial \varepsilon_{n\mathbf{k}}^{\text{FAN}}}{\partial n_{m\mathbf{q}}} + \frac{\partial \varepsilon_{n\mathbf{k}}^{\text{DDW}_{\text{RIA}}}}{\partial n_{m\mathbf{q}}} 
  \end{equation} 
 with 
\begin{multline}\label{denk_dnmq_FAN}
\frac{\partial \varepsilon_{n\mathbf{k}}^{\text{FAN}}}{\partial n_{m\mathbf{q}}} =
\frac{1}{2\omega_{m\mathbf{q}}} \sum_{\substack{\kappa\alpha\\ \kappa'\gamma}} \mathcal{F}_{\substack{\kappa\alpha\\ \kappa'\gamma}}(\mathbf{q})U_{m,\kappa'\gamma}^*(\mathbf{q})U_{m,\kappa\alpha}(\mathbf{q})
\end{multline}
\begin{multline}\label{denk_dnmq_DDW_RIA}
\frac{\partial \varepsilon_{n\mathbf{k}}^{\text{DDW}_{\text{RIA}}}}{\partial n_{m\mathbf{q}}} = \frac{-1}{4\omega_{m\mathbf{q}}}\sum_{\substack{\kappa\alpha\\ \kappa'\gamma}}   \mathcal{F}_{\substack{\kappa\alpha\\ \kappa'\gamma}}(\mathbf{\Gamma}) \\
  \Big( U_{m,\kappa\gamma}^*(\mathbf{q})U_{m,\kappa\alpha}(\mathbf{q}) + U_{m,\kappa'\gamma}^*(\mathbf{q})U_{\kappa'\alpha}^m(\mathbf{q})\Big),
\end{multline}
where the DDW$_{\text{RIA}}$ actually originates from the second-order contribution of displacing the same atom,
although it assumes a Fan-like form. We can see that within RIA all DW-type terms disappear making the calculations easier to perform (only first-order derivative of the Hamiltonian have to be computed).

All Fan-like contributions can be derived from DFPT, as explained in the appendix~\ref{perturbationth}.
In practice, in the AHC theory as formulated by Giustino \textit{ et al}\cite{Giustino2010}, 
$\mathcal{F}_{\substack{\kappa\alpha\\ \kappa'\gamma}} (\mathbf{\mathbf{q}})$ is obtained using
 Eq.~\eqref{Fan} and 
\begin{multline}\label{equationsumoverstate}
\Big\langle \frac{\partial u_{n\mathbf{k}}}{\partial R_{\kappa\alpha}(\mathbf{q})} \Big| \frac{\partial \hat{H}_{\mathbf{k,k}}}{\partial R_{\kappa'\gamma}(\mathbf{q}) }\Big| u_{n\mathbf{k}}^{(0)}\Big\rangle
=\sideset{}{'}\sum_{n'=1}^{\infty} \\
\frac{\Big\langle u_{n\mathbf{k}}^{(0)}\Big| \frac{\partial \hat{H}_{\mathbf{k,k}}}{\partial R_{\kappa\alpha}(-\mathbf{q})}  \Big| u_{n'\mathbf{k+q}}^{(0)} \Big\rangle\Big\langle u_{n'\mathbf{k+q}}^{(0)}\Big| \frac{\partial \hat{H}_{\mathbf{k,k}}}{\partial R_{\kappa'\gamma}(\mathbf{q})}  \Big| u_{n\mathbf{k}}^{(0)} \Big\rangle}{\varepsilon_{n\mathbf{k}}^{(0)}-\varepsilon_{n'\mathbf{k+q}}^{(0)}},
\end{multline}
where the infinite sum over bands is truncated in numerical calculations, while the prime after the sum symbol indicates that 
the terms with a vanishing denominator (such situation always occurs at $\Gamma$ when $n=n'$) have to be excluded.

In the resulting expression, one needs to sum over a large number of empty bands since the first-order wavefunction is expressed in the sum-over-states form. As shown by Sternheimer\cite{Sternheimer1954}, the summation over highly energetic bands can be replaced by the solution of a linear equation. This equation can then be solved iteratively with the same techniques as the ones of the DFPT approach used to calculate phonon eigenvectors and eigenenergies\cite{Baroni1987,Baroni2001,Gonze1997,Gonze2005}.
The resulting expression for the first-order wavefunction is detailed in appendix~\ref{perturbationth} of this paper (Eq.~\eqref{ch2.15}) and leads to an alternative form of Eq.~\eqref{equationsumoverstate}, proposed in Ref.~\onlinecite{Gonze2011}
\begin{multline}\label{matrix_element}
\Big\langle \frac{\partial u_{n\mathbf{k}}}{\partial R_{\kappa\alpha}(\mathbf{q})} \Big| \frac{\partial \hat{H}_{\mathbf{k,k}}}{\partial R_{\kappa'\gamma}(\mathbf{q}) }\Big| u_{n\mathbf{k}}^{(0)}\Big\rangle
=\\
\sideset{}{'}\sum_{n'=1}^{M} 
\frac{
\Big\langle u_{n\mathbf{k}}^{(0)}\Big| \frac{\partial \hat{H}_{\mathbf{k,k}}}{\partial R_{\kappa\alpha}(-\mathbf{q})}  \Big| u_{n'\mathbf{k+q}}^{(0)} \Big\rangle
\Big\langle u_{n'\mathbf{k+q}}^{(0)}\Big| \frac{\partial \hat{H}_{\mathbf{k,k}}}{\partial R_{\kappa'\gamma}(\mathbf{q})}  \Big| u_{n\mathbf{k}}^{(0)} \Big\rangle}
{\varepsilon_{n\mathbf{k}}^{(0)}-\varepsilon_{n'\mathbf{k+q}}^{(0)}}
\\
+
\Big\langle P_{c\mathbf{k+q}}{\frac{\partial u_{n\mathbf{k}}}{\partial R_{\kappa\alpha}(\mathbf{q})}}\Big| \frac{\partial \hat{H}_{\mathbf{k,k}}}{\partial R_{\kappa'\gamma}(\mathbf{q})}  \Big| u_{n\mathbf{k}}^{(0)} \Big\rangle
\end{multline}

This formulation is independent of the value of M, which is usually taken to be slightly larger than the number of bands
for which the electron-phonon renormalization is sought. It removes the cumbersome sum over states, and results in a significant speed up of the calculations\cite{Gonze2011} as well as the elimination of the convergence study on the truncation of the sum.
The complete expression for the change of eigenenergies due to electron-phonon interactions,
in the RIA, is obtained from the combination of Eqs.~\eqref{Denk(T)}, \eqref{Fan}, \eqref{Eqrewritednotfinal} and \eqref{matrix_element}.

\subsection{Beyond the rigid-ion approximation}

One can actually analyze the full expression for the derivative of the eigenenergies with respect to phonon occupation numbers, Eq.~\eqref{denk_dnmq_abbrev}, and split it into the sum of the two following contributions
\begin{equation}
\frac{\partial \varepsilon_{n\mathbf{k}}}{\partial n_{m\mathbf{q}}} = \frac{\partial \varepsilon_{n\mathbf{k}}^{\text{FAN}}}{\partial n_{m\mathbf{q}}} +\frac{\partial \varepsilon_{n\mathbf{k}}^{\text{DW}}}{\partial n_{m\mathbf{q}}},
\end{equation}
where the first term has already been identified and with
\begin{equation}
\frac{\partial \varepsilon_{n\mathbf{k}}^{\text{DW}}}{\partial n_{m\mathbf{q}}}= \frac{\partial \varepsilon_{n\mathbf{k}}^{\text{DDW}}}{\partial n_{m\mathbf{q}}}  + \frac{\partial \varepsilon_{n\mathbf{k}}^{\text{NDDW}}}{\partial n_{m\mathbf{q}}},
\end{equation}
where the DW term has been divided into a DDW and a non-diagonal DW contribution (NDDW) term defined as simply the diagonal and non-diagonal in $\kappa,\kappa'$ parts of the full DW term of Eq.~\eqref{denk_dnmq_abbrev}
\begin{multline}\label{DDWterm}
\frac{\partial \varepsilon_{n\mathbf{k}}^{\text{DDW}}}{\partial n_{m\mathbf{q}}} =\\
 \frac{1}{2\omega_{m\mathbf{q}}}\sum_{\substack{\kappa\alpha\\ \kappa'\gamma}} \delta_{\kappa'\kappa}\mathcal{D}_{\substack{\kappa\alpha\\ \kappa'\gamma}}(\mathbf{q})U_{m,\kappa'\gamma}^*(\mathbf{q})U_{m,\kappa\alpha}(\mathbf{q}) 
\end{multline}\begin{multline}\label{NDDW}
\frac{\partial \varepsilon_{n\mathbf{k}}^{\text{NDDW}}}{\partial n_{m\mathbf{q}}}
 =\\
  \frac{1}{2\omega_{m\mathbf{q}}}\sum_{\substack{\kappa\alpha\\ \kappa'\gamma}}
 (1-\delta_{\kappa\kappa'})\mathcal{D}_{\substack{\kappa\alpha\\ \kappa'\gamma}}(\mathbf{q})
    U_{m,\kappa'\gamma}^*(\mathbf{q})U_{m,\kappa\alpha}(\mathbf{q}),
\end{multline}
where the NDDW term is the mixed derivative of two different atoms in the same cell as well as its replicas.

Finally, the DDW term can be divided even further into a diagonal rigid-ion approximation
contribution ($\text{DDW}_{\text{RIA}}$) already identified and a diagonal non-rigid-ion approximation contribution ($\text{DDW}_{\text{NRIA}}$) 
\begin{equation}
\frac{\partial \varepsilon_{n\mathbf{k}}^{\text{DDW}}}{\partial n_{m\mathbf{q}}}= \frac{\partial \varepsilon_{n\mathbf{k}}^{\text{DDW}_{\text{RIA}}}}{\partial n_{m\mathbf{q}}}  + \frac{\partial \varepsilon_{n\mathbf{k}}^{\text{DDW}_{\text{NRIA}}}}{\partial n_{m\mathbf{q}}},
\end{equation}
where the $\text{DDW}_{\text{NRIA}}$ term can be obtained as the remaining components from Eq.~\eqref{denk_dnmq_RI}
\begin{multline}\label{DDWNRIA}
\frac{\partial \varepsilon_{n\mathbf{k}}^{\text{DDW}_{\text{NRIA}}}}{\partial n_{m\mathbf{q}}}
= \frac{1}{2\omega_{m\mathbf{q}}}\sum_{\substack{\kappa\alpha\\ \kappa'\gamma}}\bigg[
    \Big[ \delta_{\kappa\kappa'}\mathcal{D}_{\substack{\kappa\alpha\\ \kappa'\gamma}}(\mathbf{q})
          - \mathcal{D}_{\substack{\kappa\alpha\\ \kappa'\gamma}}(\mathbf{\Gamma}) \Big] \\
    \frac{1}{2}\Big( U_{m,\kappa\gamma}^*(\mathbf{q})U_{m,\kappa\alpha}(\mathbf{q})
                   + U_{m,\kappa'\gamma}^*(\mathbf{q})U_{m,\kappa'\alpha}(\mathbf{q})\Big) \bigg], 
\end{multline}
where the DDW$_{\text{NRIA}}$ contains only the second-order derivative of the Hamiltonian with respect to two coordinates belonging to different periodic replicas of the same atom.  
Interestingly, although this term and the DDW$_{\text{RIA}}$ originate from the diagonal Debye-Waller term, its expression includes a non-diagonal contribution. 
We argue that this definition of the diagonal terms is mathematically more convenient as this definitions holds outside of the RIA. An alternative definition, previously used in the literature, defines the DDW as the diagonal term in $\kappa,\kappa'$ at $\mathbf{q}=\mathbf{\Gamma}$ which contain the same derivative taken twice with respect to the same atom. 

We may note that our FAN and $\text{DDW}_{\text{RIA}}$ terms are the same as Eq.~16 and Eq.~15 of Ref.~\onlinecite{Gonze2011}, co-authored by some of us. Eq.~22 of the latter reference regroups our $\text{DDW}_{\text{NRIA}}$ and NDDW terms into a term that was called non-diagonal Debye-Waller.  
Incidentally, this paper studied isolated molecules, for which the only wavevector to be considered
is $\mathbf{q}=\mathbf{\Gamma}$. Thus the $\text{DDW}_{\text{NRIA}}$ contribution always vanished in that case.
We believe that the updated definitions are more general since our DDW term is really diagonal in $\kappa,\kappa'$, not only
in the isolated molecule case, but also in the periodic case, while our NDDW term is purely off-diagonal, as implied by their names and in contrast to the definition of Ref.~\onlinecite{Gonze2011}.

\section{Temperature dependence from finite differences over atomic displacements}
\label{finite_differences}

The temperature dependence can also be computed through a FP approach where, in a supercell, a set of self-consistent first-principles calculations are done with atoms displaced slightly from their equilibrium positions. 
The change of force due to atomic displacement allows one to construct the IFC, from which the phonon frequencies and eigenvectors can be deduced. 

In a DFT approach (GW behaves similarly, with an electronic self-energy replacing the DFT exchange-correlation potential), the Hamiltonian is the sum of the kinetic energy operator $\hat{T}$ and the Kohn-Sham potential $\hat V_{\text{KS}}[\rho]$. 
The latter can further be split in the sum of potentials generated by each ion $V_{l\kappa}(\mathbf{\hat r}-\mathbf{R}_{l\kappa})$ and the Hartree and exchange-correlation (Hxc) potential $\hat V_{\text{Hxc}}[\rho]$ generated by the electronic density of the system
\begin{equation}
\hat{H} = \hat{T} + \hat V_{\text{KS}}[\rho] =  \hat{T} + \sum_{l\kappa} V_{l\kappa}(\mathbf{\hat r}-\mathbf{R}_{l\kappa}) + \hat V_{\text{Hxc}}[\rho]. 
\end{equation}

Thus, the change of the Hamiltonian is related to the change of the density only through $\hat V_{\text{Hxc}}[\rho]$. 
The change of the density due to the displacement of one atom being in general affected by the displacement of another atom, the second order derivative of the Hamiltonian with respect to the displacement of two different atoms will contain contributions from $\hat V_{\text{Hxc}}[\rho]$, but not from $\hat T$ and $V_{l\kappa}(\mathbf{\hat r}-\mathbf{R}_{l\kappa})$ since they have the form of Eq.~\eqref{rigid-ion_H}. 
This can be seen as a consequence of the fact that, unlike the bare ionic potential, the Hxc potential is screened.

We can use the dielectric function $\hat \epsilon$ to describe this effect more rigorously. 
The change of Kohn-Sham potential can then be expressed as the change of ionic (bare) potential, screened by the inverse dielectric function $\hat \epsilon^{-1}$
\begin{equation}
\frac{\partial \hat V_{\text{KS}}}{\partial \mathbf{R}_{l\kappa}} = \hat \epsilon^{-1} \frac{\partial}{\partial \mathbf{R}_{l\kappa}} \sum_{l'\kappa'} \hat V_{l'\kappa'}.
\end{equation}
As the dielectric function depends on all the atomic positions, frozen-phonon DFT calculations do include non-diagonal terms.
Since, in contrast, the AHC formalism (and thus DFPT) neglects non-diagonal contributions through the RIA, it becomes interesting to devise a scheme that allows one to obtain these contributions from FP calculations.
One can then assess the validity of the AHC formalism and obtain the magnitude of the contributions neglected by the RIA.

We start from Eq.~\eqref{denk_dnmq} and compute the second order derivative of the eigenenergies through a FP approach based on second-order finite differences 
\begin{align}\label{finitediff}
\frac{\partial \varepsilon_{n\mathbf{k}}}{\partial n_{m\mathbf{q}}} =& 
\frac{1}{2 \omega_{m\mathbf{q}}} \sum_{\substack{l\kappa\alpha\\ l'\kappa'\gamma}}\frac{\partial^2 \varepsilon_{n\mathbf{k}}}{\partial R_{l\kappa\alpha}\partial R_{l'\kappa'\gamma}} \nonumber \\
  &e^{-i\mathbf{q}\cdot (\mathbf{R}_{l}-\mathbf{R}_{l'})}U_{m,\kappa'\gamma}^{*}(\mathbf{q})U_{m,\kappa\alpha}(\mathbf{q}) \nonumber \\
  =& \frac{1}{2 \omega_{m\mathbf{q}}} \frac{\partial^2}{\partial h^2} \varepsilon_{n\mathbf{k}}\Big[\Big\{   R_{l\kappa\alpha} = R_{l\kappa\alpha}^{(0)}  \nonumber\\
  &+ \left. h U_{m,\kappa\alpha}(\mathbf{q})e^{-i\mathbf{q}\cdot \mathbf{R}_{l}}  \Big\} \Big]\right|_{h=0},
\end{align}
where $h$ sets the amplitude of the FP displacement along the normal mode $(m, \mathbf{q})$, $R_{l\kappa\alpha}^{(0)}$ are the equilibrium position of the atoms, the eigenvalues $\varepsilon_{n\mathbf{k}}$ are evaluated with the atoms displaced along the phonon mode $(m, \mathbf{q})$ and the $l$, $\kappa$ and $\alpha$ indices are iterated upon within the $\{\}$. 

We also compute, from a FP approach, the first NRIA contribution to the eigenenergies renormalization, that is, the DDW$_{\rm NRIA}$ contribution (Eq.~\eqref{DDWNRIA})
\begin{widetext}
\begin{multline}\label{DDWNRIAterm}
\frac{\partial \varepsilon_{n\mathbf{k}}^{\text{DDW}_{\text{NRIA}}}}{\partial n_{m\mathbf{q}}} = \frac{1}{2\omega_{m\mathbf{q}}}\bigg[ \sum_{\kappa} \frac{\partial^2}{\partial h_{\kappa}^2}  \left.\left\langle u_{n\mathbf{k}}^{(0)}\middle| \hat{H}_{\mathbf{k,k}} \Big[\Big\{   R_{l\kappa'\alpha} = R^{(0)}_{l\kappa'\alpha}+  h_{\kappa'} U_{m,\kappa'\alpha}(\mathbf{q})e^{-i\mathbf{q}\cdot \mathbf{R}_{l}} \Big\} \Big] \middle| u_{n\mathbf{k}}^{(0)}\right\rangle\right|_{h=0}   \\
- \sum_\kappa \frac{\partial^2}{\partial h_{\kappa}^2} \left.\left\langle u_{n\mathbf{k}}^{(0)}\middle| \hat{H}_{\mathbf{k,k}} \Big[\Big\{   R_{l\kappa'\alpha} = R^{(0)}_{l\kappa'\alpha} + h_{\kappa'} U_{m,\kappa'\alpha}(\mathbf{q})\Big\} \Big] \middle| u_{n\mathbf{k}}^{(0)}\right\rangle \right|_{h=0} \\
-\sum_\gamma \frac{\partial^2}{\partial s \partial t}\left\langle u_{n\mathbf{k}}^{(0)}\middle| \hat{V}_{\text{Hxc}}\Big[\Big\{ R_{l\kappa'\alpha} = R_{l\kappa'\alpha}^{(0)}\left. + s U_{m,\kappa'\gamma}^{*}(\mathbf{q}) U_{m,\kappa'\alpha}(\mathbf{q}) + t \delta_{\alpha\gamma}\Big\} \Big] \middle| u_{n\mathbf{k}}^{(0)}\right\rangle \right|_{h=0}
 \\
 +\sum_{\kappa} \frac{\partial^2}{\partial h_{\kappa}^2}  \left\langle u_{n\mathbf{k}}^{(0)}\middle| \hat{V}_{\text{Hxc}}\left[\Big\{   R_{l\kappa'\alpha} = R^{(0)}_{l\kappa'\alpha}  +\left. h_{\kappa'} U_{m,\kappa'\alpha}(\mathbf{q}) \Big\} \right] \middle| u_{n\mathbf{k}}^{(0)}\right\rangle\right|_{h=0}\bigg],
\end{multline}
\end{widetext}
where $\{h_\kappa\}$, $s$ and $t$ are scalars introduced for finite differences purposes.
The first term computed through finite differences is a collective displacement of the atoms in a primitive cell.
The second order derivative of this term is non-zero because the curvature of the displaced potential is evaluated at fixed equilibrium ion position through the unperturbed periodic part  of the wavefunction $u_{n\mathbf{k}}^{(0)}$.

We finally proceed to calculate from a FP approach the second NRIA contribution to the renormalization, that is, the NDDW contribution
\begin{multline}\label{nddwterm}
\frac{\partial \varepsilon_{n\mathbf{k}}^{\text{NDDW}}}{\partial n_{m\mathbf{q}}} = \frac{1}{2\omega_{m\mathbf{q}}}\bigg[ \sum_{\substack{\kappa\kappa'\\ (\kappa\neq \kappa')}} \frac{\partial^2}{\partial h_{\kappa}\partial h_{\kappa'}}\\
\left\langle u_{n\mathbf{k}}^{(0)}\middle| \hat{V}_{\text{Hxc}}\Big[\Big\{ \right. R_{l\kappa''\alpha} =    R_{l\kappa''\alpha}^{(0)} +\\
 \left. \left. h_{\kappa''} U_{m,\kappa''\alpha}(\mathbf{q})e^{-i\mathbf{q}\cdot \mathbf{R}_{l}}  \Big\} \Big] \middle| u_{n\mathbf{k}}^{(0)}\right\rangle \right|_{h=0}\bigg].
\end{multline}

Antonius {\it et al} \cite{Antonius2014} computed the temperature dependence of diamond using finite differences (Eq.~\eqref{finitediff}) in the many-body $GW$ framework, which led to an additional 200 meV ZPR of the diamond bandgap with respect to the AHC value, closer to the experimental bandgap.

\section{Comparison between AHC/DFPT and finite-difference results}
\label{Calculation_Result}

In this section, we compare AHC and and FP results for some chosen $\mathbf{q}$-wavevector contributions to the ZPR of the diamond bandgap. 
We study the $\mathbf{q}$-wavevector contributions to the ZPR instead of the full ZPR since the converged $\mathbf{q}$-point integration required by the full ZPR is computationally out of reach for FP (it requires 70x70x70 $\mathbf{q}$-point grids and associated supercells). 
For the full $\mathbf{q}$-point integration within AHC, please see Ref.~\onlinecite{Ponce2014}.      

The calculation of structural properties in this work are based on Density Functional Theory (DFT) \cite{Martin2004} within the local density approximation (LDA) \cite{Ceperley1980,Perdew1981}.
We use a planewave basis set to represent the wavefunctions and account for the core-valence interaction using norm-conserving pseudopotentials \cite{Troullier1991}. 
The valence electrons of carbon treated explicitly in our ab-initio calculations are 2s$^{2}$2p$^{2}$.
All the calculations are done using the ABINIT software package\cite{Gonze2009}.

Convergences studies with a tolerance of 0.5mHa per atom on the total energy led to the use of a 6x6x6 $\Gamma$-centered Monkhorst-Pack sampling \cite{Monkhorst1976} of the Brillouin zone and an energy cut-off of 30 Hartree for the truncation of the planewave basis set. 
The lattice parameter of 6.675 Bohr was obtained by structural optimization.


The phonon frequencies were calculated using Eq.~\eqref{ch2.6} for the DFPT method and a second-order derivative of the total energy with respect to atomic displacements converged with a Richardson interpolation of order 4 for the finite-difference method.
Also, for the latter method and the $\mathbf{q}=L$ point, a 2x2x2 supercell with a 3x3x3 $\Gamma$-centered Monkhorst-Pack sampling were used to ensure the $\mathbf{k}$-point sampling remained equivalent to the one used for the DFPT calculation on the primitive cell.    
The comparison between FP and DFPT for the phonon frequencies is given in Table~\ref{Comparison_freq}.
The discrepancies between the two methods remain within 0.2\% for the two $\mathbf{q}$-points considered, which demonstrates both the equivalence of the two methods and the convergence of our calculations. 
\begin{table}
\caption{\label{Comparison_freq}Comparison of phonon frequencies computed from FP and from DFPT.}
\begin{ruledtabular}
\begin{tabular}{|c|c|c|c|c|}
 $\mathbf{q}$-point  & Mode & $\omega$ DFPT [Ha] & $\omega$ FP [Ha] & diff. [\%]\\
\hline
 $\mathbf{\Gamma}$  & LO+TO & 0.00606474  & 0.00605226 & 0.2058 \\ 
 $\mathbf{L}$       & TA    & 0.00250256  & 0.00250111 & 0.0580 \\
           & LA    & 0.00494158  & 0.00494344 & 0.0376 \\
           & TO    & 0.00564786  & 0.00565244 & 0.0810 \\
           & LO    & 0.00577597  & 0.00577016 & 0.1007 \\
\end{tabular}
\end{ruledtabular}
\end{table}

We now assess the accuracy of our AHC implementation and quantify the impact of the RIA. 
The contribution of the same two $\mathbf{q}$-points to the ZPR at $\mathbf{k}=\mathbf{\Gamma}$ and $\mathbf{k}=\mathbf{L}$ is given in Table \ref{Comparison}. This table presents the renormalization of the different eigenenergies due to electron-phonon coupling for the 4 first distinguishable bands of diamonds. The $\mathbf{\Gamma_{25'}}$  valence band maximum is three-fold degenerate as well as the $\mathbf{\Gamma_{15}}$ conduction band. In diamond, the conduction band minimum is located between the $\mathbf{k}=\mathbf{\Gamma}$ and $\mathbf{k}=\mathbf{X}$ points. At the $\mathbf{k}=\mathbf{L}$, the valence $\mathbf{L_1}$ and conduction $\mathbf{L_{3'}}$ bands are only doubly degenerate. It can also be noted that our results show the contribution of some $\mathbf{q}$-points to the ZPR with values in very close agreements (within 2~meV) with Figure 1 of Ref.~\onlinecite{Antonius2014} that show the electron-phonon coupling energies (there is therefore a conversion factor of 1/2). 

The AHC results are split into Fan and DDW$_\text{RIA}$ contributions that are computed using Eq.~\eqref{denk_dnmq_FAN} and Eq.~\eqref{denk_dnmq_DDW_RIA}, respectively. 

The importance of the rigid-ion approximation (RIA) can be deduced by computing the DDW$_\text{NRIA}$ and NDDW  terms through finite difference calculations using Eq. \eqref{DDWNRIAterm} and Eq. \eqref{nddwterm}, respectively. 
It can be noticed that the DDW$_\text{NRIA}$ and the NDDW terms are equal at $\mathbf{q}=\mathbf{\Gamma}$, due to the $T_{2g}$ optical mode of the diamond crystal for which $U_{m,\kappa\alpha}=-U_{m,\kappa'\alpha}$ when $\kappa\neq\kappa'$, as expected from Eqs.~\eqref{DDWNRIA} and \eqref{NDDW}.

The impact of the RIA remains below 23\% and is usually much smaller for wavevectors other than the zone-center one. 
To assess completely the validity of the approximation, we should do a full $\mathbf{q}$-point integration on the BZ.
However, since the ZPR converges extremely slowly with the number of $\mathbf{q}$-points~\cite{Ponce2014}, this would require huge supercell calculations, which are currently computationally out of reach.    

Finally, the sum of the AHC and NRIA contributions can be compared to the finite difference calculations done using Eq.~\eqref{finitediff}. All the FP calculations are also converged with a Richardson interpolation of order 4.
The discrepancies remain below 7~$\mu$eV in absolute value. 
Such discrepancies can be attributed to numerical noise or anharmonicity.
Indeed, the finite displacements selected in the FP method ensure a good compromise between these two sources of error, so that both are present, but as small as allowed by our convergence criteria.
\begin{table*}
\caption{\label{Comparison}Comparison between the AHC/DFPT and FP results for the contributions of specific $\mathbf{q}$-points to the ZPR of several bands of diamond. The column ``diff." is the sum of the AHC and the NRIA contributions minus the FP result. }
\begin{tabular}{|c|c| S[table-format=5.4]| S[table-format=4.4]|S[table-format=5.4]| S[table-format=1.4]| S[table-format=3.4]| S[table-format=2.2]|  S[table-format=5.4]| S[table-format=5.4]| S[table-format=3.3]|  }
\hline
\hline
  $\mathbf{q}$-point &  $\mathbf{k}$-point  &\multicolumn{3}{c|}{AHC [meV]}  & \multicolumn{3}{c|}{NRIA  [meV]} & {sum all } & {FP} & { diff. }\\
\cline{0-7}
                                 &                   &  {Fan} & {DDW$_\text{RIA}$} & sum AHC &  {DDW$_\text{NRIA}$} & {NDDW} & {\% NRIA } & {[meV] }  & {[meV] } & {[$\mu$eV] }   \\
\hline
	  $\mathbf{\Gamma}$ &  $\mathbf{\Gamma_1}$  &  -32.8174 & 20.2868  &  -12.5306 &  0.7250  & 0.7250 & 13.09 & -11.0806  & -11.0809  & 0.308 \\  
         &  $\mathbf{\Gamma_{25'}}$    & -332.4265 & 357.2564 &   24.8300 &  1.7991  & 1.7991 & 12.66 &  28.4282  &  28.4289  &-0.638 \\   
         &  $\mathbf{\Gamma_{15}}$     & -330.4363 & 316.2021 &  -14.2342 &  0.1924  & 0.1924 &  2.78 & -13.8494  & -13.8497  & 0.338 \\ 
         & $\mathbf{\Gamma_{2'}}$      &  -63.3087 &  32.3760 &  -30.9327 &  0.1499  & 0.1499 &  0.98 & -30.6328  & -30.6335  & 0.748 \\   
\cline{2-10}           
         & $\mathbf{L_{2'}}$           &  -67.8598 &  46.8783 &  -20.9814 &  1.1410  & 1.1410 & 12.20 & -18.6993  & -18.6999  & 0.554 \\   
         &  $\mathbf{L_1}$             & -146.0806 & 129.4769 &  -16.6037 &  0.5663  & 0.5663 &  7.32 & -15.4710  & -15.4714  & 0.367 \\
         & $\mathbf{L_{3'}}$           & -311.2306 & 321.3287 &   10.0981 &  1.4805  & 1.4805 & 22.67 &  13.0590  &  13.0592  &-0.307 \\   
         &  $\mathbf{L_3}$             & -473.0115 & 292.4656 & -180.5458 &  0.0779  & 0.0779 &  0.09 & -180.3900 &-180.3937  & 6.977 \\
\hline\hline         
    $\mathbf{L}$  &$\mathbf{\Gamma_1}$          & -116.4278 &  62.6966 &  -53.7312 & 3.2318 & -2.3250 &  1.72 & -52.8256  & -52.8245  & 1.104 \\
         & $\mathbf{\Gamma_{25'}}$     & -922.8240 &1104.1052 &  181.2812 & 6.0491 & -3.7542 &  1.25 & 183.5761  & 183.5771  &-1.017 \\
         & $\mathbf{\Gamma_{15}}$      &-1250.8082 & 977.2263 & -273.5819 & 1.0840 & -2.0900&  0.37 & -274.5878  &-274.5881  & 0.244 \\
         & $\mathbf{\Gamma_{2'}}$      & -407.6022 & 100.0584 & -307.5438 & 0.3437 & -2.1920 &  0.60 & -309.3921  &-309.3973  & 5.131 \\
\cline{2-10}            
         & $\mathbf{L_{2'}}$           & -234.2353 & 144.8781 &  -89.3572 & 4.3856 & -2.8996 &  1.69 & -87.8712  & -87.8728  & 1.542 \\
         &  $\mathbf{L_1}$             & -620.7070 & 400.1500 & -220.5570 & 2.6651 & -2.1582 &  0.23 & -220.0501  &-220.0525  & 2.359 \\
         &  $\mathbf{L_{3'}}$          &-1018.9788 & 993.0698 &  -25.9090 & 5.1401 & -3.3984 &  7.21 & -24.1674  & -24.1672  &-0.210 \\
         &  $\mathbf{L_3}$             & -740.6821 & 903.8683 &  163.1862 & 0.7991 & -1.8919 &  0.67 & 162.0934  & 162.0935  &-0.142 \\
\hline
\hline
\end{tabular}
\end{table*}


\section{Conclusions}
\label{Conclusions}
The renormalization of eigenenergies due to electron-phonon coupling can be computed by different methods. In this paper, we have reviewed three of them: the first-principle molecular dynamics method, the frozen-phonon (FP) method and the Allen-Heine-Cardona (AHC) method based on density-functional perturbation theory.  
The two first methods are equivalent within the adiabatic harmonic approximation while the third is only equivalent when the rigid-ion approximation (RIA) is also performed.

The theory's key ingredient is the second-order derivative of the eigenenergies with respect to two atomic displacements. This derivative gives rise to a term stemming from the first-order modification of the wavefunction called the Fan term and a term corresponding to a second-order perturbation of the Hamiltonian called the Debye-Waller (DW) term.
Although the two terms were discovered separately in the 50's, there was a lot of confusion in the litterature until Baumann realised in 1974 the complementarity of these two terms for the computation of the zero-point motion renormalization (ZPR).   

The present paper compared in detail two (AHC/DFPT and FP) of the three approaches still used today to calculate the ZPR. We considered the first for its efficiency in the computation of the ZPR at arbitrary $\mathbf{q}$-points, crucial for periodic system, and the later, to go beyond the RIA and study its impact on the calculated ZPR.   

Also, in this paper, we derived Brook's theorem in the first-principle context and obtained an expression for the eigenenergy renormalization from the finite temperature extension of Janak's theorem (see Eqs.~\eqref{equallity} and \eqref{ch2.7bis}).

We also rederived how, within the RIA, the translational invariance (Eq.~\eqref{ch2.18bis}) allows to express the DW contribution in terms of first-order derivatives of the Hamiltonian only. 

A major contribution of this paper is the clarification of the terms appearing beyond the RIA made in the AHC theory. The DW term is divided into a diagonal rigid-ion approximation contribution ($\text{DDW}_{\text{RIA}}$), a diagonal non-rigid-ion approximation  contribution ($\text{DDW}_{\text{NRIA}}$), and a non-diagonal DW contribution (NDDW). This allows all the term's definition to be coherent with their respective names (see Eqs.~\eqref{denk_dnmq_FAN}, \eqref{denk_dnmq_DDW_RIA}, \eqref{DDWNRIA} and \eqref{NDDW}). 

Nevertheless, due to the computational limitation related to the evaluation of the second-order derivative of the Hamiltonian, the NRIA terms cannot be computed with DFPT. 
Therefore we derived the equations related to the FP approach, which allows us to numerically evaluate the NRIA terms in the case of bulk diamond (see Eqs.~\eqref{finitediff}, \eqref{DDWNRIAterm} and \eqref{nddwterm}). 

For the diamond phonon frequencies, the discrepancy between the FP and AHC approaches is below 0.2\%. The differences of ZPR between the two methods are always below 7$\mu eV$ in absolute value, which strengthens our confidence in the theory and numerical implementation within the ABINIT software. 
These small differences are attributed to the unavoidable numerical noise and anharmonicity in the FP calculations. 
The impact of the RIA is also evaluated for two $\mathbf{q}$-points ($\mathbf{\Gamma}$ and $\mathbf{L}$) and is found to be as large as 23\% for the $\Gamma$ zone center contribution to the ZPR but is usually much smaller for other $\mathbf{q}$-points contributions.

\section{Acknowledgements}
\label{Acknowledgements}
This work was supported by the FRS-FNRS through a FRIA grant (S.P.) and a FNRS grant (Y.G.). Moreover, A. M. would like to acknowledge financial support from the Futuro in Ricerca grant No.~RBFR12SW0J of the Italian Ministry of Education, University and Research.
The authors would like to thank Yann Pouillon and Jean-Michel Beuken for their valuable technical support and help with the test and build systems of ABINIT.
Computational ressources have been provided by the supercomputing facilities of the Universit\'e catholique de Louvain (CISM/UCL) and the Consortium des \'Equipements de Calcul Intensif en F\'ed\'eration Wallonie Bruxelles (CECI) funded by the Fonds de la Recherche Scientifique de Belgique (FRS-FNRS) under Grant No.~2.5020.11.

\appendix
\section{Technicalities}
\label{Technicalities}

\subsection{Integrals of phonon wavefunctions and powers of the position operator}
\label{integral}

We evaluate the integrals present in Eq.~\eqref{eqdeptempisole}, with the Taylor expansion of the eigenenergy.

In the harmonic approximation, the phonon wavefunctions are obtained by solving the Schr\"odinger equation for the harmonic oscillator.
We obtain\cite{Griffiths1995}
\begin{equation}\label{phononwfhermite}
\chi_{s_m,m}(z) = \left(\frac{\omega_m}{\pi}\right)^{1/4} \frac{H_{s_m}(\xi)}{\sqrt{2^{s_m} s_m!}}e^{-\frac{\xi^2}{2}},
\end{equation}
where $H_{s_m}(\xi)=(-1)^{s_m} e^{\xi^2}\frac{d^{s_m}}{d\xi^{s_m}}e^{-\xi^2}$ is the Hermite polynomial and $\xi=\sqrt{\omega_m}z$ is a dimensionless position variable. 
Hermite polynomials satisfy the following orthonormality condition
\begin{equation}
\int H_p(\xi)H_{s_m}(\xi)e^{-\xi^2}d\xi = s_m! \sqrt{\pi}2^{s_m} \delta_{ps_m}.
\end{equation}
The phonon wavefunctions are thus normalized, with 
\begin{equation}
\int \chi_{s_m,m}(\xi)^*\chi_{s_m,m}(\xi)d\xi = 1.
\end{equation}
The first-order integral cancels out, as an odd function integrates to zero.
\begin{equation}
\int \chi_{s_m,m}(z)^*z\chi_{s_m,m}(z)dz= 0.
\end{equation}
 
Finally, the square of the dimensionless position operator in second quantization can be expressed as
\begin{equation}\label{secondquantiz}
\xi^2 = \frac{1}{2}(a+a^+)^2. 
\end{equation}
giving
\begin{equation}
\begin{split}
&\langle 0|a^{s_m} \frac{1}{2}(a+a^+)^2 (a^{+})^{s_m}|0\rangle \\
&= \langle 0|a^{s_m} \frac{1}{2}(aa^++a^+a) (a^{+})^{s_m}|0\rangle \\
&= \langle 0|a^{s_m} \frac{1}{2}(2aa^+ -1) (a^{+})^{s_m}|0\rangle \\
&= \langle 0|a^{s_m+1} (a^{+})^{s_m+1}|0\rangle - \frac{1}{2}\langle 0|a^{s_m} (a^{+})^{s_m}|0\rangle   \\
&= (s_m+1)! - \frac{1}{2}s_m!= s_m!\left( \frac{2s_m+1}{2}\right).
\end{split}
\end{equation}
Hence, the second-order in the Taylor expansion yields
\begin{equation}
\int \chi_{s_m,m}(z)^*z^2\chi_{s_m,m}(z)dz = \frac{2s_m+1}{2 \omega_m}.
\end{equation}

\subsection{Convention for the unperturbed periodic system}
\label{conventionfor}
Following the same convention as Gonze\cite{Gonze1997}, the unperturbed wavefunction can be obtained as the product of a phase factor and a periodic function (Bloch's theorem)
\begin{equation}\label{periodicwf}
\Psi_{n\mathbf{k}}^{(0)}(\mathbf{r}) = (N_{BvK}\Omega_0)^{-1/2}e^{i\mathbf{k}\cdot\mathbf{r}}u_{n\mathbf{k}}^{(0)}(\mathbf{r}),
\end{equation}
where $N_{BvK}$ is the number of unit cells repeated in the Born-von Karman periodic box, $\Omega_0$ the volume of the unperturbed unit cell, $n$ the band index and $\mathbf{k}$ label the wave vector of the wavefunction. 

The periodic part of the Bloch wavefunction, in Eq.~\eqref{periodicwf}, is subject to the following orthonormalization condition
\begin{equation}
\left\langle u_{n'\mathbf{k}}^{(0)}\middle| u_{n\mathbf{k}}^{(0)} \right\rangle = \delta_{n'n},
\end{equation}
where the scalar product of periodic functions is defined as
\begin{equation}
\left\langle f \middle| g \right\rangle  =  \frac{1}{\Omega_0}\int_{\Omega_0} f^*(\mathbf{r})g(\mathbf{r})d\mathbf{r}.
\end{equation}

For a generic operator we follow the following convention
\begin{equation}\label{def_operator}
O_{\mathbf{k,k'}} = e^{-i\mathbf{k}\cdot\mathbf{r}}Oe^{-i\mathbf{k'}\cdot\mathbf{r'}}.
\end{equation}

\subsection{Perturbation theory}
\label{perturbationth}

The perturbation theory for a periodic system is based on the idea that the solution of a reference system (usually the equilibrium ground state one-body Schr\"odinger equation) is known.
In general, the perturbation can be incommensurate with the periodic system and characterized by a wave vector $\mathbf{q}$.
If the amplitude of the perturbation is characterized by a small scalar parameter $\lambda$, then any observable $X(\lambda)$ can be expressed as a power serie
\begin{multline}
 X(\lambda) = X^{(0)}+(\lambda X_{\mathbf{q}}^{(1)}+\lambda^{*} X_{\mathbf{-q}}^{(1)}) +(\lambda^2 X_{\mathbf{q,q}}^{(2)} \\
 + \lambda\lambda^{*}X_{\mathbf{q,-q}}^{(2)}+ \lambda^{*}\lambda X_{\mathbf{-q,q}}^{(2)}+\lambda^{*2} X_{\mathbf{-q,-q}}^{(2)})+ \cdots,
\end{multline}
were we use the superscript notation as a shorthand for derivatives: $X_{\mathbf{q}}^{(i)}=\left. d^i X_{\mathbf{q}}/id\lambda^i \right|_{\lambda=0}$. 
The perturbed Schr\"odinger equation now depends explicitely on the parameter $\lambda$
\begin{equation}\label{ch2.8}
 H(\lambda)\left|\Psi_{n\mathbf{k}}(\lambda)\right\rangle = \varepsilon_{n\mathbf{k}}(\lambda)\left|\Psi_{n\mathbf{k}}(\lambda)\right\rangle,
\end{equation}
where $n$ is the band index, $\mathbf{k}$ is a wavevector in the Brillouin zone. 
The solutions $\left|\Psi_{n\mathbf{k}}(\lambda)\right\rangle$ must fulfill the normalization condition $\left\langle\Psi_{n\mathbf{k}}(\lambda)\middle|\Psi_{n\mathbf{k}}(\lambda)\right\rangle=1$. 
The Hamiltonian $\hat{H}(\lambda)$ depends parametrically on the atomic position $\textbf{R}_{l\kappa}$ of the atom $\kappa$ in the cell $l$. 

The translated first-order wave function becomes
\begin{equation}
\Psi_{n\mathbf{k,q}}^{(1)}(\mathbf{r+R}_a)=e^{i(\mathbf{k+q})\cdot \mathbf{R}_a}\Psi_{n\mathbf{k,q}}^{(1)}(\mathbf{r}),
\end{equation}
where $\mathbf{R}_a$ is a vector of the real space lattice. 
We can then factorize out the phase factor to map the incommensurate problem into a problem commensirate with periodicity of the unperturbed one. 
To this end, we introduce the periodic first-order wave functions
\begin{equation}
 u_{n\mathbf{k,q}}^{(1)} = (N_{BvK}\Omega_0)^{1/2}e^{-i(\mathbf{k+q})\cdot\mathbf{r}} \Psi_{n\mathbf{k,q}}^{(1)}(\mathbf{r}).
\end{equation} 
We define $S_{M\mathbf{k+q}}$ as the space of the $M$ low-lying $\mathbf{k+q}$ states, 
$M$ being larger or equal to the index of the highest band for which 
we aim to compute the temperature-dependent behaviour. 
We define $P_{M\mathbf{k+q}}$ as the projector on $S_{M\mathbf{k+q}}$.
We also define $P_{c\mathbf{k+q}}$ as the projector on the subspace of $\mathbf{k+q}$ ground-state wavefunctions  complementary to $S_{M\mathbf{k+q}}$. 
The first-order derivatives of wavefunctions can be split in two contributions, one that is contained inside 
$S_{M\mathbf{k+q}}$ and one that belongs to its complimentary,
\begin{equation}\label{decomp_u1}
\left|u_{n\mathbf{k,q}}^{(1)}\right\rangle = 
\left|P_{M\mathbf{k+q}}u_{n\mathbf{k,q}}^{(1)}\right\rangle
+
\left|P_{c\mathbf{k+q}}u_{n\mathbf{k,q}}^{(1)}\right\rangle.
\end{equation}
$\left|P_{M\mathbf{k+q}}u_{n\mathbf{k,q}}^{(1)}\right\rangle$ can be easily computed using standard perturbation theory,
\begin{multline}\label{perturbeq_M}
\left|P_{M\mathbf{k+q}}u_{n\mathbf{k,q}}^{(1)}\right\rangle =\\
- \sum_{n'}^{M} \frac{\left|u_{n'\mathbf{k+q}}^{(0)}\right\rangle\left\langle u_{n'\mathbf{k+q}}^{(0)} \right| H_{\mathbf{k+q,k}}^{(1)}\left| u_{n\mathbf{k}}^{(0)} \right\rangle}{\varepsilon_{n'\mathbf{k+q}}^{(0)}-\varepsilon_{n\mathbf{k}}^{(0)}}.
\end{multline}
For $\left|P_{c,\mathbf{k+q}}u_{n\mathbf{k,q}}^{(1)}\right\rangle$, we want to avoid the summation over
an infinite number of states. So, in the spirit of DFPT, we minimize
\begin{multline}
E_{\mathbf{-q,q}}^{(2)+}\{u^{(0)},u^{(1)}\} = E_{\mathbf{-q,q}}^{(2)}\{u^{(0)},u^{(1)}\}\\
-\sum_{nn'}^{MM}\Lambda_{nn'\mathbf{k,k+q}} \left\langle u_{n'\mathbf{k+q}}^{(0)}\middle|
P_{c\mathbf{k+q}}u_{n\mathbf{k,q}}^{(1)}\right\rangle +c.c.,
\end{multline}
where $E_{\mathbf{-q,q}}^{(2)}\{u^{(0)},u^{(1)}\}$ is given in Eq. 42 of Ref.~\onlinecite{Gonze1997},
under the constraint of parallel-transport gauge\cite{Gonze1995},
\begin{equation}\label{transportgauge}
\left\langle u_{n'\mathbf{k+q}}^{(0)}
\middle|
P_{c\mathbf{k+q}}u_{n\mathbf{k,q}}^{(1)}
\right\rangle = 0,
\end{equation}
with the Lagrange parameters
\begin{equation}
\Lambda_{nn'\mathbf{k,k+q}}^* = \Lambda_{n'n\mathbf{k+q,k}}.
\end{equation}
The minimum of this expression with respect to variations of $P_{c\mathbf{k+q}}u_{n\mathbf{k,q}}^{(1)}$
leads to the canonical Euler-Lagrange equation
\begin{multline}\label{sternheimerlike}
\bigg( H_{\mathbf{k+q,k+q}}^{(0)}-\varepsilon_{n\mathbf{k}}^{(0)}  \bigg)\Big| 
P_{c\mathbf{k+q}}u_{n\mathbf{k,q}}^{(1)}\Big\rangle =\\
 -  H_{\mathbf{k+q,k}}^{(1)}\Big| u_{n\mathbf{k}}^{(0)}\Big\rangle  + \sum_{n'}\Lambda_{nn'\mathbf{k,k+q}}^*\Big|u_{n'\mathbf{k+q}}^{(0)}\Big\rangle.
\end{multline}

We can then pre-multiply on each side by $\Big\langle u_{n''\mathbf{k+q}}^{(0)}\Big|$ where $n''\in [1,M]$ and get 
\begin{multline}
\Big( \varepsilon_{n''\mathbf{k+q}}^{(0)}-\varepsilon_{n\mathbf{k}}^{(0)} \Big) \overbrace{\Big\langle u_{n''\mathbf{k+q}}^{(0)}\Big| u_{n\mathbf{k,q}}^{(1)}\Big\rangle}^{=0 \text{ due to Eq. \eqref{transportgauge}}} \\
+ \Big\langle u_{n''\mathbf{k+q}}^{(0)}\Big| H_{\mathbf{k+q,k}}^{(1)}\Big| u_{n\mathbf{k}}^{(0)}\Big\rangle \\
= \sum_{n'}\Lambda_{nn'\mathbf{k,k+q}}^* \underbrace{\Big\langle u_{n''\mathbf{k+q}}^{(0)}\Big|u_{n'\mathbf{k+q}}^{(0)}\Big\rangle}_{\delta_{n''n'}}.
\end{multline}
This leads to the equation
\begin{equation}\label{HFtheorem}
\Lambda_{nn''\mathbf{k+q,k}}^* = \Big\langle u_{n''\mathbf{k+q}}^{(0)}\Big|H_{\mathbf{k+q,k}}^{(1)}\Big| u_{n\mathbf{k}}^{(0)}\Big\rangle.
\end{equation}
We can then substitute Eq.~\eqref{HFtheorem} inside Eq.~\eqref{sternheimerlike} and get
\begin{multline}
\Big( H_{\mathbf{k+q,k+q}}^{(0)}-\varepsilon_{n\mathbf{k,q}}^{(0)}  \Big)\Big| 
P_{c\mathbf{k+q}}u_{n\mathbf{k,q}}^{(1)}\Big\rangle =\\
 - \Big(\underbrace{1-\sum_{n'}^{N_{\text{max}}}\Big|u_{n'\mathbf{k+q}}^{(0)}\Big\rangle \Big\langle u_{n'\mathbf{k+q}}^{(0)}\Big|}_{P_{c\mathbf{k+q}}} \Big) H_{\mathbf{k+q,k}}^{(1)} \Big|u_{n\mathbf{k}}^{(0)}\Big\rangle.
\end{multline}
and finally \begin{multline}\label{ch2.15}
 P_{c,\mathbf{k}+\mathbf{q}}\left( H_{\mathbf{k+q,k+q}}^{(0)}-\varepsilon_{n\mathbf{k}}^{(0)}\right)P_{c,\mathbf{k+q}}\left|u_{n\mathbf{k,q}}^{(1)}\right\rangle\\
  =-P_{c,\mathbf{k+q}}H_{\mathbf{k+q,k}}^{(1)}\left|u_{n\mathbf{k}}^{(0)}\right\rangle.
\end{multline}

Supposing $H^{(1)}$ has already be determined by the usual DFPT self-consistency loop over occupied states only (in which $M=N_{val}$), the $u_{n\mathbf{k,q}}^{(1)}$ are found by solving Eq.~\eqref{ch2.15}, combined with Eqs.~\eqref{decomp_u1} and \eqref{perturbeq_M}.  

\bibliography{article_v25}

\end{document}